\documentclass[useAMS,onecolumn,usenatbib,usegraphicx]{mn2e}
\usepackage{graphicx}
\input{epsf}

\newif\ifAMStwofonts
\AMStwofontstrue

\newcommand{\go}{\mathrel{\raise.3ex\hbox{$>$}\mkern-14mu
             \lower0.6ex\hbox{$\sim$}}}
\newcommand{\lo}{\mathrel{\raise.3ex\hbox{$<$}\mkern-14mu
             \lower0.6ex\hbox{$\sim$}}}
\newcommand{\lp}{\left(}
\newcommand{\rp}{\right)}
\newcommand{\lb}{\left[}
\newcommand{\rb}{\right]}

\newcommand{\me}{m_{\rm e}}

\newcommand{\alphaf}{\alpha_{\rm F}}
\newcommand{\vecE}{{\bmath E}}
\newcommand{\vecB}{{\bmath B}}
\newcommand{\vechatB}{\hat{\bmath B}}
\newcommand{\veck}{{\bmath k}}
\newcommand{\vechatk}{\hat{\bmath k}}

\newcommand{\vechaty}{\hat{\bmath y}}

\newcommand{\vechatz}{\hat{\bmath z}}
\newcommand{\vechatzp}{\hat{\bmath z}'}

\newcommand{\vecr}{{\bmath r}}

\newcommand{\Bq}{B_{\rm Q}}
\newcommand{\thetab}{\theta_B}

\newcommand{\vecD}{{\bmath D}}
\newcommand{\vecepspl}{\bepsilon^{\rm (p)}}
\newcommand{\vecepsvp}{\Delta\bepsilon^{\rm (v)}}

\newcommand{\epsvp}{\Delta\epsilon^{\rm (v)}}
\newcommand{\vpa}{a}
\newcommand{\vpatwo}{\hat{a}}
\newcommand{\vpq}{q}
\newcommand{\vpm}{m}
\newcommand{\vpr}{r}
\newcommand{\vprtwo}{\hat{r}}
\newcommand{\deltavp}{\delta_{\rm V}}

\newcommand{\de}{{\rm d}}
\newcommand{\vecv}{\bmath V}
\newcommand{\dvecJ}{\bmath{\delta J}}
\newcommand{\dv}{\bmath{\delta V}}
\newcommand{\dE}{\bmath{\delta E}}
\newcommand{\dB}{\bmath{\delta B}}

\newcommand{\dns}{\delta N_s}

\newcommand{\singh}{\sin\theta_{B}}
\newcommand{\singhsq}{\sin^2\theta_{B}}
\newcommand{\cosgh}{\cos\theta_{B}}
\newcommand{\cosghsq}{\cos^2\theta_{B}}
\newcommand{\ggtheta}{\lp1-n\beta\cosgh\rp}
\newcommand{\ggthetaplus}{\lp1+n\beta\cosgh\rp}
\newcommand{\ggthetas}{\lp1-n\beta_s\cosgh\rp}
\newcommand{\ggthetanon}{\lp1-\beta\cosgh\rp}
\newcommand{\zetagh}{\lp\cosgh-\zeta\singh\rp}

\newcommand{\NGJ}{N_{\rm GJ}}

\newcommand{\wpl}{\omega_{p}}
\newcommand{\wpe}{\omega_{p-}}
\newcommand{\wpp}{\omega_{p+}}
\newcommand{\wc}{\omega_{c}}
\newcommand{\wce}{\omega_{c-}}
\newcommand{\wcp}{\omega_{c+}}

\newcommand{\polarb}{\beta_{\rm p}}
\newcommand{\polarbpl}{\beta_0}
\newcommand{\polarbvp}{\beta_{\rm V}}

\newcommand{\veceps}{{\bepsilon}}
\newcommand{\vecone}{{\bf I}}
\newcommand{\vecsigma}{{\bsigma}}
\newcommand{\vecx}{{\bf x}}

\newcommand{\vecz}{{\bf z}}

\newcommand{\nuG}{\nu_{\rm 1}}
\newcommand{\Ps}{P_{\rm 1}}

\title[]{Wave Modes in the Magnetospheres of Pulsars and Magnetars}
\author[C. Wang and D. Lai]{Chen Wang$^{1,2}$ and Dong Lai$^{2,1}$ \\
$^{1}$National Astronomical Observatories, Chinese Academy of Sciences.
A20 Datun Road, Chaoyang District, Beijing 100012, China \\
$^{2}$Center for Radiophysics and Space Research, 
Department of Astronomy, Cornell University.
Ithaca, NY 14853, USA \\
{\rm E-mail: wangchen@bao.ac.cn, dong@astro.cornell.edu}}
\date{Accepted 2006 xxx,
      Received 2006 xxx;
      in original form 2006 xxx}

\pagerange{\pageref{firstpage}--\pageref{lastpage}}

\begin{document}

\maketitle

\label{firstpage}

\begin{abstract}
We study the wave propagation modes in the relativistic streaming pair
plasma of the magnetospheres of pulsars and magnetars, focusing on the
effect of vacuum polarization. We show that the combined plasma and
vacuum polarization effects give rise to a vacuum resonance, where
``avoided mode crossing'' occurs between the extraordinary mode and
the (superluminous) ordinary mode. When a photon propagates from the
vacuum-polarization-dominated region at small radii to the
plasma-dominated region at large radii, its polarization state may
undergo significant change across the vacuum resonance. We map out the
parameter regimes (e.g., field strength, plasma density and Lorentz
factor) under which the vacuum resonance occurs and examine how wave
propagation is affected by the resonance. Some possible applications
of our results are discussed, including high-frequency radio emission
from pulsars and possibly magnetars, and optical/IR emission from
neutron star surfaces and inner magnetospheres.
\end{abstract}

\begin{keywords}
plasmas -- polarization -- waves -- star: magnetic fields -- pulsars:
general
\end{keywords}

\setcounter{equation}{0}
\section{Introduction} \label{sec:intro}

The magnetospheres of pulsars and magnetars consist of relativistic
electron-positron pair plasmas, plus possibly a small amount of
ions. These plasmas can affect the radiation produced in the inner
region of the magnetosphere or the stellar surface.  Understanding the
property of wave propagation in pulsar/magnetar magnetospheres is
important for the interpretation of various observations of these
objects.

Radio emission from pulsars (at least ``normal''--non millisecond --
pulsars) likely originates from close to the stellar surface, within a
few percent of light cylinder radius (e.g., Blaskiewicz et al.~1991,
Kramer et al.~1997).  A number of studies have been devoted to the
propagation effect of radio waves in pulsar magnetospheres (e.g.,
Cheng \& Ruderman 1979; Barnard \& Arons 1986; Barnard 1986;
Lyubarskii \& Petrova 1998; Melrose \& Luo 2004; Petrova 2006).  Some
of the observed polarization properties of pulsar emission, such as
orthogonal modes (in which the polarization position angle exhibits a
sudden $\sim 90^\circ$ jumps; e.g. Stinebring et al.~1984a, 1984b) and
circular polarization (e.g., Radhakrishnan \& Rankin 1990; Han et
al.~1998; You \& Han 2006) may be explained by the propagation effect.
In addition, optical/IR radiation may be produced in the inner
magnetosphere or surface of magnetized neutron stars. For example,
while for most radio pulsars the optical and near IR flux is thought
to be dominated by magnetospheric emission, several middle-aged
pulsars (PSR B0656+14, PSR B0950+08 and Geminga) also exhibit a
surface optical component (e.g., Mignani, de Luca \& Caraveo 2004;
Kargaltsev et al.~2005, Mignani et al. 2006).  The optical emission
detected in several thermally emitting, isolated neutron stars mostly
likely has a surface origin (e.g., Kaplan et al.~2003; Haberl et
al.~2004; Haberl 2005; van KerKwijk \& Kaplan 2006).  Finally, the
optical/IR emission detected from a number of magnetars may originate
from a hot corona near the stellar surface (Beloborodov \& Thompson
2006).

Wave modes in pulsar magnetospheres have been studied in a number of
papers under different assumptions about the plasma composition and
the velocity distribution of electron-position pairs (e.g., Melrose \&
Stoneham 1977; Arons \& Barnard 1986; Lyutikov 1998; Melrose et
al.~1999; Asseo \& Riazuelo 2000). In this paper, we reinvestigate the
property of wave propagation in the magnetospheres of pulsars and
magnetars, focusing on the competition between the plasma effect and
the effect of vacuum polarization.  It is well known that in the
strong magnetic field typically found on a neutron star, the
electromagnetic dispersion relation is dominated by vacuum
polarization (a prediction of quantum electrodynamics; e.g.,
Heisenberg \& Euler 1936; Adler 1971; see Schubert 2000 for extensive
bibliography) at high photon frequencies (e.g. X-rays) and by the
plasma effect at sufficiently low frequencies (e.g., radio waves). But
where is the ``boundary'' at which the two effects are ``equal'' and
what are the mode proportion in the ``boundary'' regime? These are the
questions we are trying to address in this paper.  We show that the
combined plasma and vacuum polarization effects give rise to a {\it
vacuum resonance}: For a given plasma parameters and external magnetic
field, there exists a special photon frequency at which the plasma
effect and vacuum polarization effects ``cancel'' each other.  A more
physical way to describe the resonance is as follows: Consider a
photon propagating in the inhomogeneous pulsar/magnetar magnetosphere
with varying plasma density (and/or distribution function) and
magnetic field.  For certain parameter regimes of the photon
frequency, plasma density and magnetic field strength --- to be
determined in the following sections, the photon may traverse from the
vacuum-polarization-dominated region to the plasma-dominated region or
vice versa. This transition point (location) is the vacuum
resonance. When the photon crosses this resonance, its polarization
state may undergo significant change.  The goal of our paper is to map
out the parameter regimes under which the vacuum resonance may occur
and to elucidate how wave propagation may be affected by the
resonance.

Vacuum resonance in cold, non-streaming plasmas have been studied
before (e.g., Gnedin et al. 1978; M\'{e}sz\'{a}ros \& Veutura 1979;
Lai \& Ho 2002, 2003a). In the atmospheres of highly magnetized
neutron stars, the resonance can significantly affect the surface
emission spectrum and polarization (Ho \& Lai 2003; Lai \& Ho 2003a,
b; van Adelsberg \& Lai 2006). We note that while some previous papers
on wave modes in pulsar magnetospheres (e.g. Arons \& Barnard 1986)
did include the vacuum polarization contributions to the dielectric
tensor,
the vacuum resonance phenomenon was neglected because it is
unimportant at the low frequencies and magnetic fields
they considered.

Our paper is organized as follows. In \S 2, we give the expression for
the dielectric tensor of a relativistic pair plasma charactering the
magnetosphere of pulsars/magnetars, including the contribution due to
vacuum polarization.  In \S 3, we derive the general expression for
wave modes in the combined ``plasma + vacuum'' medium, and show that
the vacuum resonance arises for a wide range of magnetosphere
parameters. In \S 4 we study the evolution of wave mode across the
vacuum resonance.  In most of this paper, we consider cold, streaming
plasma with a single Lorentz factor $\gamma$. We examine the effect of
a more general $\gamma$ distribution in \S 5 and the case of
opposite plasma streams in \S 6. In \S 7 we discuss possible
applications of our results.

\section{Dielectric tensor for an streaming electron-positron plasma}
\label{sec:dielectric}

We consider an electron-positron plasma in the magnetosphere of a
neutron star (NS). Let $N_-$, $N_+$ be the number densities of
electrons and positrons, respectively, $N=N_-+N_+$ the total density,
and $f=N_+/N$ the positron fraction. The corotation region of the
magnetosphere is usually assumed to have Goldreich-Julian charge
density
\begin{equation}
\rho_e=-\frac{1}{2\pi c}{\bf\Omega}\cdot\vecB,
\end{equation}
where ${\bf\Omega}$ is the angular velocity of the star. 
This is not necessarily
the case in the open-field line region. In this paper we shall use the 
Goldreich-Julian number density as a fiducial value:
\begin{equation}
\NGJ=\frac{\Omega B}{2\pi ec}
    \simeq 7.0\times 10^{10} B_{12}\Ps^{-1}\,{\rm cm}^{-3}.
    \label{eq:NGJ}
\end{equation}
where $B_{12}=B/(10^{12}\,{\rm G})$, $\Ps$ is the spin period in units
of 1$\,$s. The actual particle density $N$ is larger than $\NGJ$ by a
factor of $\eta$, i.e. $N=\eta\NGJ$, with $\eta \geq 1$. If the charge
density is equal to the Goldreich-Julian value, then $\eta(1-2f)=1$,
but we will not entirely restrict ourselves to this constraint.

Although the plasma is expected to be relativistic, it is useful to
define the (nonrelativistic) cyclotron frequencies and plasma
frequencies of electron and position:
\begin{eqnarray}
\wc    &=&\wcp=\wce=\frac{eB}{m_ec},\\
\wpe^2 &=&\frac{4\pi N_- e^2}{m_e}=(1-f)\wpl^2, \\
\wpp^2 &=&\frac{4\pi N_+ e^2}{m_e}=f\wpl^2, \\
\wpl^2 &=&\frac{4\pi N e^2}{m_e}.
\end{eqnarray}
The characteristic values are
\begin{equation}
\nu_{\rm c} = \frac{\wc}{2\pi} = 2.795\times10^9\,B_{12}\mbox{ GHz} \label{eq:nuc}
\end{equation}
\begin{equation}
\nu_{\rm p} = \frac{\wpl}{2\pi} = 8.960\times10^3 N^{1/2}\,{\rm Hz} 
= 2.370\,\eta^{1/2}B_{12}^{1/2}\Ps^{-1/2}\,{\rm GHz}, \label{eq:nup}
\end{equation}

Because of the very short cyclotron/synchrotron decay time of
elections and positions ($\approx
3\times10^{-16}B_{12}^{-2}\gamma\,{\rm sec}$) , all the particles in
the magnetosphere quickly lose their transverse momenta and stay in
the lowest Landau level. Thus the magnetosphere pair plasma can be
considered as one-dimensional, with the particles streaming along the
field line. The Lorentz factor $\gamma$ of the streaming motion is
uncertain. In the polar-cap region of a pulsar, primary particles may
be accelerated to very high energy ($\gamma\sim10^6-10^7$) by a
field-aligned electric field. The bulk of the plasma produced in an
electromagnetic cascade may have lower energies,
$\gamma\sim10^2-10^4$, with multiplicity factor $\eta\sim10^2-10^5$
(e.g., Dangherty \& Harding 1982; Hibschman \& Arons 2001). Kunzl et
al. (1998) argued that too high a density of secondary particles in
the magnetosphere is in contradiction to the observed low-frequency
emission from radio pulsars, implying $\eta\lo 100$.  The physical
parameters for the plasma in the closed-field-line region of a pulsar
are also not well constrained.  It was suggested that a pair plasma
density larger than $N_{GJ}$ may be present, maintained by conversion
of $\gamma$-rays from the pulsar's polar-cap and/or out-gap
accelerators (see Wang et al.~1998; Ruderman 2003).

For magnetars, recent theoretical work suggests that a corona
consisting mainly of relativistic pairs with $\gamma\sim 10^3$ (and a
wide spread in $\gamma$) may be generated by crustal magnetic field
twisting/shearing due to starquakes (Thompson et al.~2002; Beloborodov
\& Thompson 2006). The plasma density is of order $N\sim
|\nabla\times{\bf B}|/(4\pi e)\sim B/(4\pi er)$ (for a twist angle of
order unity), implying $\eta=N/N_{GJ}\sim c/(2\Omega r)\simeq 2\times
10^3(R/r)$ (where $R$ is the stellar radius). There are roughly equal
amount of electrons and positrons, $f\simeq1/2$, with the electrons and
positrons streams in opposite directions.

\subsection{Cold, Streaming Pair Plasma}

We first consider a cold electron-positron plasma with all the
electron streaming with velocity $\bmath V_{-,0}$ and positron all with
$\bmath V_{+,0}$ which is also along the magnetic filed $\bmath B_0$. The
dielectric tensor for such a plasma was derived by Melrose \& Stoneham
(1977) based on Lorentz transformation (see also Melrose 1973). Here
we outline a derivation based on classical magneto-ionic theory.

The equation of motion of a given charge species (mass $m_s=m_e$, charge
$q_s=\pm e$, $s=\pm$) reads
\begin{equation}
\frac{\de}{\de t} \lp\gamma_s m_s \vecv_s\rp= q_s\vecE + \frac{q_s}{c}\vecv_s\times\vecB,
\label{eq:emotion}
\end{equation}
where $\beta_s=V_s/c$, $\gamma_s=(1-V_s^2/c^2)^{-1/2}$,
$\vecv_s=\vecv_{s,0}+\dv_s$, $\vecE=\dE$, $\vecB=\vecB_0+\dB$. Here
$\dv_s$, $\dE$, $\dB$ are associated with the disturbance in the
plasma, and have the form $e^{i(\veck\cdot\vecr-\omega t)}$. With
$\de\dv_s/\de t = -i\omega\dv_s+ic\lp\veck\cdot\vecv_{s,0}\rp\dv_s $
and $\dB=(c\veck/\omega)\times\dE$, we can solve for $\dv_s$ in terms
of the components of $\dE$. The current density associated with the
disturbance is
\begin{eqnarray}
\dvecJ &=& \sum_s \lp N_s q_s\dv_s+ \dns q_s\vecv_{s,0}\rp, \label{eq:J-v1} 
\end{eqnarray}
where the sum runs over each charged particle species $s$ (electron
$e$ and position $p$). The density perturbation $\dns$ is
determined by the continuity equation, $\partial N_s/\partial t =-
\nabla\cdot\lp N_s\vecv_{s,0}\rp$, which gives
$\dns=N_s(\veck\cdot\dv_s)/(\omega-\veck\cdot\vecv_{s,0})$.  The
conductivity tensor is defined by $\dvecJ=\bsigma\cdot\dE$, and the
dielectric tensor is given by $\veceps=\vecone+i(4\pi
c/\omega)\vecsigma$, where $\vecone$ is the unit tensor.
In the coordinate system $x'y'z'$ with ${\hat\vecB_0}/|\vecB_0|$
along $\vecz'$, and $\veck$ in the $\vecx'$-$\vecz'$ plane, such that
$\vechatk\times\vechatB_0=-\sin\thetab\vechaty'$ ($\thetab$ is the
angle between $\vechatk$ and $\vechatB_0$), we find
\begin{equation}
\vecepspl
= \lb \begin{array}{ccc}
S  & iD & A \\
-iD & S & -iC  \\
A & iC & P
\end{array} \rb, \label{eq:diel1} 
\end{equation}
with
\begin{eqnarray}
    S &=& 1+\sum_s f_{s,11}, \nonumber\\
    D &=& \sum_s f_{s,12}, \nonumber\\
    A &=& \sum_s \zeta_s f_{s,11}, \nonumber\\
    C &=& \sum_s \zeta_s f_{s,12}, \nonumber\\
    P &=& 1+\sum_s\lp f_{s,\eta}+\zeta_s^2f_{s,11}\rp, \label{eq:SDACP}
\end{eqnarray}
where
\begin{eqnarray}
f_{s,11} &=& -\frac{v_s\gamma_s^{-1}}{1-u_s\gamma_s^{-2}\ggthetas^{-2}} , \nonumber\\
f_{s,12} &=& -\frac{{\rm sign}\lp q_s\rp u_s^{1/2}v_s\gamma_s^{-2}\ggthetas^{-1}}
                 {1- u_s\gamma_s^{-2}\ggthetas^{-2}}, \nonumber\\
f_{s,\eta} &=& -\frac{v_s}{\gamma_s^3 \ggthetas^2}, \nonumber \\
\zeta_s    &=& \frac{n\beta_s\singh}{1-n\beta_s\cosgh},\label{eq:fzeta}
\end{eqnarray}
(the subscript ``0'' has been suppressed),
$n=ck/\omega$ is the refractive index and
the relevant dimensionless quantities are
\begin{equation}
u_{-}=u_{+} = u = \frac{\wc^2}{\omega^2},\label{eq:u}
\end{equation}
\begin{equation}
v_{-} = \frac{\wpe^2}{\omega^2}, \qquad
v_{+} = \frac{\wpp^2}{\omega^2}, \qquad
v = \frac{\wpl^2}{\omega^2} = v_{-}+v_{+}. \label{eq:v}
\end{equation}

\subsubsection{Pair Plasma With the Same Velocity ($\beta_-=\beta_+=\beta$)}

Consider the case where all the electrons and positrons have the same
velocity ($\vecv_-=\vecv_+$, $\beta_-=\beta_+=\beta$,
$\gamma_-=\gamma_+=\gamma$). In this paper, we will focus on the
regime where $\wc\gg \gamma\omega\ggtheta$, or
$u\gamma^{-2}\ggtheta^{-2}\gg1$, i.e. the photon frequency shifted to
the plasma rest frame is much lower than the cyclotron frequency. In
this regime, the components of dielectric tensor can be simplified to
\begin{eqnarray}
    S &=& 1+f_{11}, \nonumber\\
    D &=& f_{12}, \nonumber\\
    A &=& \zeta f_{11}, \nonumber\\
    C &=& \zeta f_{12}, \nonumber\\
    P &=& 1+ f_{\eta}+\zeta^2f_{11},
\end{eqnarray}
with
\begin{eqnarray}
f_{11}  &=& \sum_s f_{s, 11}  \simeq vu^{-1}\gamma\ggtheta^{2}, \nonumber\\
f_{12}  &=& \sum_s f_{s, 12}  \simeq -\lp1-2f\rp vu^{-1/2}\ggtheta,\nonumber\\
f_{\eta}&=& \sum_s f_{s, \eta}\simeq -v\gamma^{-3}\ggtheta^{-2}, \nonumber \\
\zeta   &=& n\beta\singh\lp1-n\beta\cosgh\rp^{-1}. \label{eq:fzetasimple}
\end{eqnarray}

\subsubsection{Pair Plasma with opposite velocity (the case of $\beta_-=-\beta_+=\beta$)}

Suppose the plasma is composed of two opposite streams: one is the
electron stream with $\beta_-=\beta$ and the other is the
positron stream with $\beta_+=-\beta$. 
For $\wc\gg \gamma\omega(1\pm n\beta\cos\theta_B)$,
Eqs.~(\ref{eq:fzeta}) simplify to:
\begin{eqnarray}
f_{+,11} &\simeq& fvu^{-1}\gamma\ggthetaplus^2, \nonumber\\
f_{-,11} &\simeq& (1-f)vu^{-1}\gamma\ggtheta^2, \nonumber\\
f_{+,12} &\simeq& fvu^{-1/2}\ggthetaplus, \nonumber\\
f_{-,12} &\simeq& -(1-f)vu^{-1/2}\ggtheta, \nonumber\\
f_{+,\eta} &\simeq& -fv\gamma^{-3}\ggthetaplus^{-2}, \nonumber\\
f_{-,\eta} &\simeq& -(1-f)v\gamma^{-3}\ggtheta^{-2}, \nonumber\\
\zeta_\pm  &=& \mp n\beta\singh\lp1\pm n\beta\cosgh\rp^{-1}.
\end{eqnarray}

\subsection{Pair Plasma with a Distribution of $\gamma$}
The pair plasma in pulsar magnetosphere many have some spread in
the Lorentz factors, although the precise distribution is unknown.
For a general distribution function $f_s(\gamma_s)$, normalized 
$\int f_s(\gamma_s)\de\gamma_s=1$, we can average our result in 
$\S$2.1 to obtain the dielectric tensor components:
\begin{eqnarray}
S  &=& 1 +\sum_s\int f_{s,11} f_s(\gamma_s)d\gamma_s , \nonumber\\
D  &=& \sum_s\int f_{s,12} f_s(\gamma_s)d\gamma_s , \nonumber\\
A  &=& \sum_s\int \zeta_s f_{s,11} f_s(\gamma_s)d\gamma_s , \nonumber\\
C  &=& \sum_s\int \zeta_s f_{s,12} f_s(\gamma_s)d\gamma_s , \nonumber\\
P  &=& 1+\sum_s\int\lp f_{s,\eta}+\zeta_s^2 f_{s,11} \rp 
           f_s(\gamma_s)d\gamma_s. \label{eq:SDACP_distr}
\end{eqnarray} 
These expressions agree with the result derived using standard kinetic
theory [e.g. Krall \& Trivelpiece 1986, equation (8.10.11); 
Arons \& Barnard~1986; Melrose \& Stoneham~1977, Lyutikov~1998].

Given the uncertainty in the $\gamma$-distribution, we shall consider the 
simplest flat distribution, for both electrons and positrons:
\begin{equation}
f(\gamma) = \left\{ \begin{array}{ll}
       1/(2\Delta\gamma) & \gamma_c-\Delta\gamma < \gamma < \gamma_c+\Delta\gamma \\
       0 & {\rm otherwise}
       \end{array} \right. \label{eq:fdistribution}
\end{equation}
where we shall assume $\gamma_{\rm min}=\gamma_c-\Delta\gamma \gg 1$
(so that $|\beta|\simeq 1$ for all particles) and
$(\gamma_c+\Delta\gamma)\omega(1\pm n\beta\cosgh) \ll \wc$ (so that
the Doppler-shifted photon frequency is always below cyclotron
resonance). Then the components of the dielectric tensor can be
evaluated analytically.

For the case of $\beta_-=\beta_+=\beta$, substituting
Eqs.~(\ref{eq:fzetasimple}) and (\ref{eq:fdistribution}) in
Eq.~(\ref{eq:SDACP_distr}), we obtain
\begin{eqnarray}
    S &=& 1+F_{11}, \nonumber\\
    D &=& F_{12}, \nonumber\\
    A &=& \zeta F_{11}, \nonumber\\
    C &=& \zeta F_{12}, \nonumber\\
    P &=& 1+ F_{\eta}+\zeta^2F_{11},
\end{eqnarray}
with
\begin{eqnarray}
F_{11} &\simeq&  vu^{-1}\gamma_c\ggtheta^2, \label{eq:f11fd}\\
F_{12} &\simeq&  -\lp1-2f\rp vu^{-1/2}\ggtheta, \label{eq:f12fd}\\
F_{\eta} &=& -v\gamma_c^{-3}\lp1-\Delta\gamma^2/\gamma_c^2\rp^{-2}
           \ggtheta^{-2}. \label{eq:fetafd}
\end{eqnarray}
Note that $F_{11}$ and $F_{12}$ are unchanged compared to the case of
delta-function distribution (Eq.~\ref{eq:fzetasimple}), while
$F_{\eta}$ is changed by a factor
$\lp1-\Delta\gamma^2/\gamma_c^2\rp^{-2}$.

For the case of $\beta_-=-\beta_+=\beta$, substituting
Eqs.~(\ref{eq:fzeta}) and (\ref{eq:fdistribution}) in
Eq.~(\ref{eq:SDACP_distr}), we obtain
\begin{eqnarray}
    S &=& 1+\sum_s F_{s,11}, \nonumber\\
    D &=& \sum_s F_{s,12}, \nonumber\\
    A &=& \sum_s \zeta_s F_{s,11}, \nonumber\\
    C &=& \sum_s \zeta_s F_{s,12}, \nonumber\\
    P &=& 1+\sum_s\lp F_{s,\eta}+\zeta_s^2F_{s,11}\rp, \label{eq:SDACP_oppdistr}
\end{eqnarray}
with 
\begin{eqnarray}
F_{+,11} &\simeq& fvu^{-1}\gamma_c\ggthetaplus^2, \label{eq:f11fd_opp_plus}\\
F_{-,11} &\simeq& (1-f)vu^{-1}\gamma_c\ggtheta^2, \label{eq:f11fd_opp_minus}\\
F_{+,12} &\simeq&  fvu^{-1/2}\ggthetaplus, \label{eq:f12fd_opp_plus}\\
F_{-,12} &\simeq&  -(1-f)vu^{-1/2}\ggtheta, \label{eq:f12fd_opp_plus}\\
F_{+,\eta} &=&  -fv\gamma_c^{-3}\lp1-\Delta\gamma^2/\gamma_c^2\rp^{-2}
           \ggthetaplus^{-2}. \label{eq:fetafd_opp_plus}\\
F_{-,\eta} &=&  -(1-f)v\gamma_c^{-3}\lp1-\Delta\gamma^2/\gamma_c^2\rp^{-2}
           \ggtheta^{-2}. \label{eq:fetafd_opp_minus}
\end{eqnarray}

\subsection{Correction due to Vacuum Polarization}

Vacuum polarization contributes a correction to the dielectric tensor:
\begin{equation}
\vecepsvp = \lp\vpa - 1\rp\vecone + \vpq\vechatB\vechatB,
\end{equation}
where $\vecone$ is the unit tensor and $\vechatB=\vecB/B$ is the unit
vector along $\vecB$ 
(here we use $\vecB$, $\vechatB$ to denote $\vecB_0$, $\vechatB_0$
for simple notations). The magnetic permeability tensor $\bmu$ also
deviates from unity because of vacuum polarization, with the inverse
permeability given by
\begin{equation}
\bmu^{-1}=\vpa\vecone+\vpm\vechatB\vechatB.
 \label{eq:permeab}
\end{equation}
In the low frequency limit $\hbar\omega\ll\me c^2$, general
expressions for the vacuum polarization coefficients $\vpa$, $\vpq$,
and $\vpm$ are given in Adler (1971) and Heyl \& Hernquist~(1997).
For $B\ll\Bq=m_e^2c^3/(e\hbar)=4.414\times 10^{13}$~G, they are given
by
\begin{equation}
\vpa  =  1-2\deltavp, \qquad
\vpq  =  7\deltavp, \qquad
\vpm  =  -4\deltavp, \label{eq:vpmlo}
\end{equation}
where
\begin{equation}
\deltavp = \frac{\alphaf}{45\pi}\lp\frac{B}{B_Q}\rp^2 
=2.650\times10^{-8}B_{12}^2 \label{eq:deltavp}
\end{equation}
and $\alphaf=e^2/\hbar c=1/137$ is the fine structure constant.  For
$B\gg\Bq$, simple expressions for $a$, $q$, $m$ are given in Ho \&
Lai~(2003) (see also Potekhin et al. 2004 for general fitting
formulae).

When $|\epsvp_{ij}|\ll 1$ or $B/B_Q\ll 3\pi/\alphaf$
($B\ll 5\times10^{16}$~G), the plasma and vacuum contributions to
the dielectric tensor can be added linearly, i.e.,
$\veceps=\vecepspl+\vecepsvp$.
In the frame with $\vechatB$ along $\vechatzp$,
\begin{eqnarray}
\lb\veceps\rb_{\vechatzp=\vechatB} 
 & = & \lb \begin{array}{ccc}
S'  & iD & A \\
-iD & S' & -iC  \\
A & iC & P'
\end{array} \rb, \label{eq:epsij}
\end{eqnarray}
with $S'=S+\vpatwo$, $P'=P+\vpatwo+\vpq$ and $\vpatwo=\vpa-1$, 

\section{Wave Modes in a Cold Streaming Plasma}
\label{sec:modes}

Here we consider the case of a pair plasma all streaming with the same
velocity $\beta$ along the field line. The effect of finite spread in
$\gamma$ will be studied in \S \ref{sec:vspread}, and the case of
opposite streams will be considered in \S 6.

\subsection{Equations for the Wave Modes}

Using the electric displacement $\vecD=\veceps\cdot\vecE$ and
equation~(\ref{eq:permeab}) in the Maxwell equations, we obtain the
equation for plane waves with $\vecE\propto
e^{i(\veck\cdot\vecr-\omega t)}$
(henceforth we use $\vecE$ to denote $\dE$, and use
$\vecB$ to denote $\vecB_0$)
\begin{equation}
\left\{ \frac{1}{\vpa}\epsilon_{ij}+n^2\lb\hat{k}_i\hat{k}_j-\delta_{ij}
 - \frac{\vpm}{\vpa}(\hat{k}\times\hat{B})_i(\hat{k}\times\hat{B})_j
 \rb \right\} E_j=0, \label{eq:nrefract}
\end{equation}
where $n=ck/\omega$ is the refractive index and $\vechatk=\veck/k$.
In the coordinate system $xyz$ with $\veck$ along the $z$-axis and
$\vecB$ in the $x$-$z$ plane, such that
$\vechatk\times\vechatB=-\sin\thetab\vechaty$, the components of
dielectric tensor are given by [compared to eq.~(\ref{eq:epsij})]
\begin{eqnarray}
\epsilon_{xx} &=& S'\cos^2\thetab - 2A\sin\thetab\cos\thetab +P'\sin^2\thetab, \nonumber \\
\epsilon_{yy} &=& S', \nonumber\\
\epsilon_{zz} &=& S'\sin^2\thetab + 2A\sin\thetab\cos\thetab +P'\cos^2\thetab, \nonumber \\
\epsilon_{xy} &=& -\epsilon_{yx} = i\lp D\cos\thetab-C\sin\thetab\rp, \nonumber \\
\epsilon_{xz} &=& \epsilon_{zx}~~= A\cos2\thetab+(S'-P')\sin\thetab\cos\thetab, \nonumber \\
\epsilon_{yz} &=& -\epsilon_{zy} = -i\lp D\sin\thetab+C\cos\thetab\rp, \label{eq:epsijx}
\end{eqnarray}
The $z$-component of equation~({\ref{eq:nrefract}) gives
\begin{equation}
E_z = -\epsilon_{zz}^{-1}\lp\epsilon_{zx}E_x+\epsilon_{zy}E_y\rp.
\label{eq:EzExEy}
\end{equation}
Reinserting this back into equation~(\ref{eq:nrefract}) yields
\begin{eqnarray}
\lp \begin{array}{cc}
\eta_{xx} - n^2 & \eta_{xy} \\
\eta_{yx} & \eta_{yy} - \vpr n^2
\end{array} \rp
\lp \begin{array}{c} E_x \\ E_y \end{array} \rp = 0, \label{eq:nrefractmatrix}
\end{eqnarray}
where $\vpr = 1+\vprtwo \equiv 1 + (\vpm/\vpa)\sin^2\thetab$ and
\begin{eqnarray}
\eta_{xx} 
& = & \frac{1}{\vpa\epsilon_{zz}}\lp\epsilon_{zz}\epsilon_{xx}
      -\epsilon_{xz}\epsilon_{zx}\rp  =  \frac{1}{\vpa\epsilon_{zz}} \lp S'P'-A^2\rp,
      \label{eta_xx}\\
\eta_{yy} 
& = & \frac{1}{\vpa\epsilon_{zz}}\lp\epsilon_{zz}\epsilon_{yy}
      -\epsilon_{yz}\epsilon_{zy}\rp \nonumber \\
& = & \frac{1}{\vpa\epsilon_{zz}} \lb(S'^2-D^2-S'P'+C^2)\sin^2\thetab+
      2(AS'-CD)\sin\thetab\cos\thetab + S'P'-C^2\rb,  \\
\eta_{yx} 
& = & -\eta_{xy} = \frac{1}{\vpa\epsilon_{zz}}
      \lp\epsilon_{zz}\epsilon_{yx}-\epsilon_{yz}\epsilon_{zx}\rp \nonumber \\
& = & \frac{-i}{\vpa\epsilon_{zz}}\lb P'D\cos\thetab -
      S'C\sin\thetab + A\lp D\sin\thetab-C\cos\thetab\rp  \rb. \label{eta_xy}
\end{eqnarray}


\subsection{$B=\infty$ limit without QED effect}
The above expressions are valid for the general dielectric tensor
(Eq.~\ref{eq:epsij}). We now consider the $B=\infty$ limit introduced by
e.g., Tsytovitch \& Kaplan (1972) and Arons \& Barnard (1986). In this
regime, the magnetic field is sufficiently large so that the cyclotron
frequency are large compared to the Lorentz-shifted wave frequency. At
the same time, $B$ is not really infinity so that the wave propagation
is dominated by plasma effect and we neglect the QED correction in the
dielectric tensor. The approximate elements of the dielectric tensor
are
\begin{equation}
f_{11}\sim vu^{-1}\gamma\sim0, \quad f_{12}\sim-(1-2f)vu^{-1/2}\sim0,
\end{equation}
\begin{equation}
f_{\eta}\simeq -v\gamma^{-3}\lp1-n\beta\cosgh\rp^{-2},
\end{equation}
\begin{equation}
S\simeq1, \quad P\simeq1+f_{\eta}, \quad D\simeq0, \quad A\simeq0, \quad C\simeq0,
\end{equation}
\begin{equation}
\epsilon_{zz}\simeq 1+f_\eta\cos^2\theta_B.
\end{equation}
Equations~(\ref{eta_xx})~--~(\ref{eta_xy}) then reduce to
\begin{eqnarray}
\eta_{xx} &\simeq& \frac{1+f_{\eta}}{1+f_{\eta}\cosghsq}, \nonumber\\
\eta_{yy} &\simeq& 1, \nonumber\\
\eta_{yx} &\simeq& -\eta_{xy} \simeq0.
\end{eqnarray}
Solving equation~(\ref{eq:nrefractmatrix}), we find (see Arons \&
Barnard 1986)
\begin{equation}
n^2=\eta_{xx}=\frac{1+f_{\eta}}{1+f_{\eta}\cosghsq}
\end{equation}
or
\begin{equation}
(\omega^2-c^2k_\parallel^2)\lb1-\frac{\omega_{\rm p}^2}
{\gamma^3\omega^2(1-\beta ck_\parallel/\omega)^2}\rb - c^2k_\perp^2=0
\end{equation}
(where $k_\parallel=k\cos\theta_B$, $k_\perp=k\sin\theta_B$) for the
ordinary mode (O-mode) and
\begin{equation}
n^2=1
\end{equation}
for the extraordinary mode (X-mode). The polarization of the O-mode is
given by
\begin{equation}
\left|\frac{E_{y}}{E_{x}}\right|=\left|\frac{\eta_{yx}}{\eta_{yy}-{\eta_xx}}\right|\sim0,
\end{equation}
\begin{equation}
\left|\frac{E_{z}}{E_{x}}\right|=\left|\frac{f_{\eta}\singh\cosgh}{1+f_{\eta}\cosghsq}\right|
=\left|\frac{n^2-1}{\tan\theta_B}\right|,
\end{equation}
and that for the X mode is
\begin{equation}
\left|\frac{E_{x}}{E_{y}}\right|\sim0, \quad \left|\frac{E_{z}}{E_{y}}\right|\sim0.
\end{equation}
Thus, the X-mode is a transverse wave with the electric field vector in the 
$\veck\times\vecB$ direction, while the O-mode is polarized in the plane spanned by $\veck$ 
and $\vecB$.

\begin{figure}
\centering
\includegraphics[height=10cm, angle=-90]{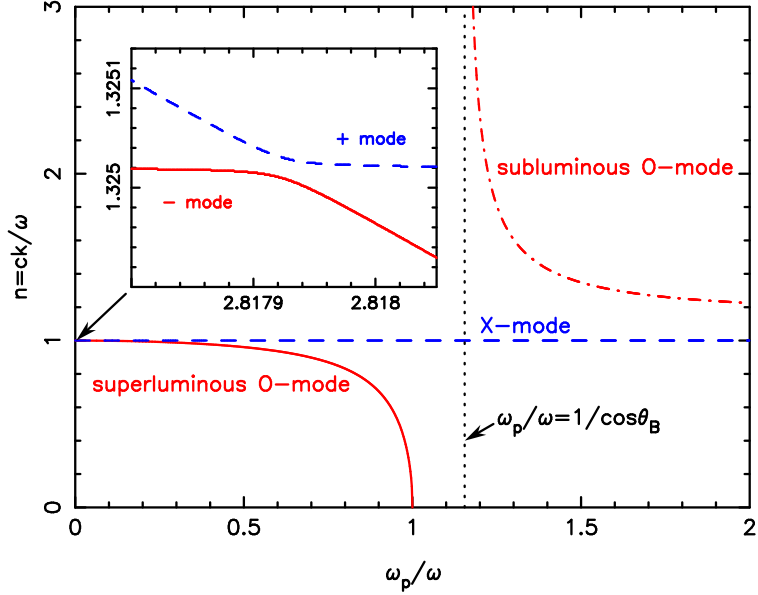}
\caption{ The refractive indices $n=ck/\omega$ of wave modes as a
function of $\omega_{\rm p}/\omega$ for a cold plasma with $\gamma=1$
and the wave propagation angle $\theta_B=30^\circ$. The solid line
shows the superluminous O-mode, the dashed line the X-mode, and the
dot-dashed line the subluminous O-mode. The insert shows the blowup of
the ``avoided crossing'' region between the X-mode and the
superluminous O-mode due to vacuum resonance. The mixed modes are
labeled ``+'' mode and ``$-$'' mode.  For the insert, the $x$-axis
gives $10^4\omega_{\rm p}/\omega$, and $y$-axis gives $10^8(n-1)$, and
the other parameters are $B=10^{12}\,$G, 
$N=N_{\rm GJ}$ with $P=1\,$s.
\label{fig:n_v} }
\end{figure}

\begin{figure}
\centering
\includegraphics[height=10cm, angle=-90]{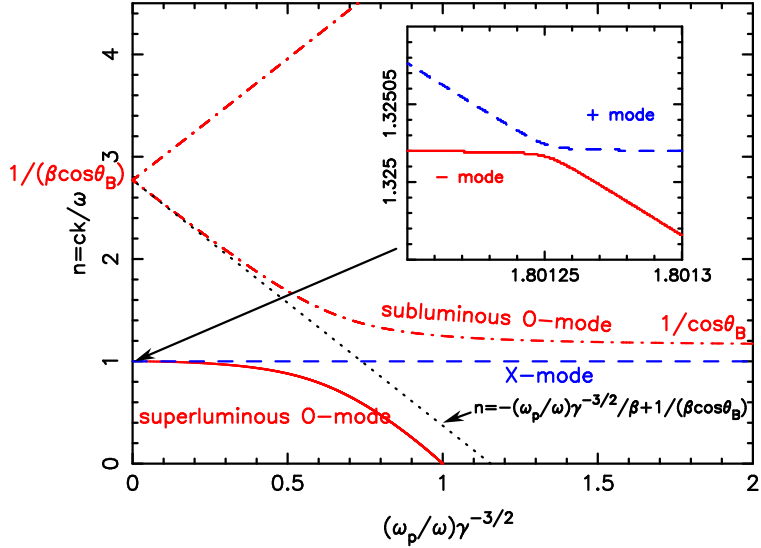}
\caption{ Same as Fig.1 except for $\gamma=1.1$, and the $x$-axis
gives $(\omega_{\rm p}/\omega)\gamma^{-3/2}$. Note that for
$(\omega_{\rm p}/\omega)\gamma^{-3/2}\ll 1$, the subluminous O-mode
has $n=-(\omega_{\rm p}/\omega)\gamma^{-3/2}+1/(\beta\cos\theta_B)$,
and it is shown as a dotted line in the figure.
\label{fig:n_v_g1.1} }
\end{figure}

\begin{figure}
\centering
\includegraphics[height=10cm, angle=-90]{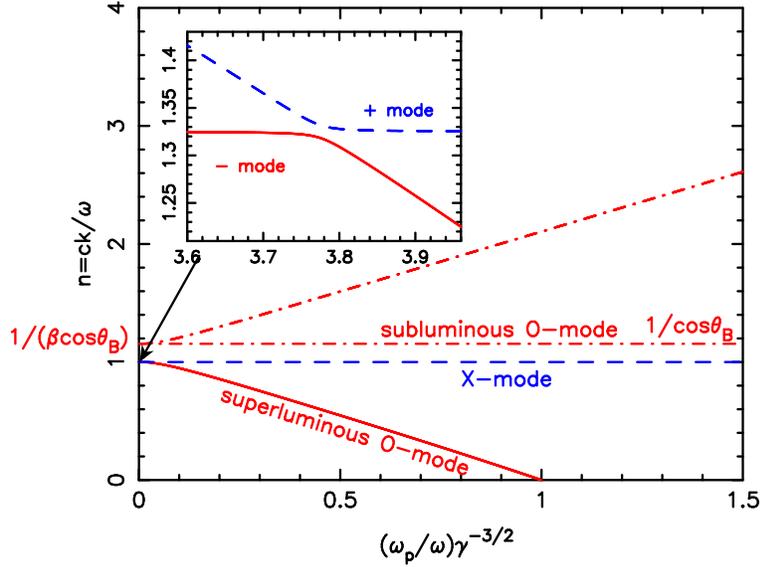}
\caption{ Same as Figure \ref{fig:n_v_g1.1} except for
$\gamma=10^3$. In the insert, the $x$-axis gives $10^5(\omega_{\rm
p}/\omega)\gamma^{-3/2}$ and the $y$-axis $10^8(n-1)$.
\label{fig:n_v_g1000} }
\end{figure}

Figures~\ref{fig:n_v}~--~\ref{fig:n_v_g1000} depict the refractive
indices $n=ck/\omega$ of different modes as a function of
$(\omega_{\rm p}/\omega)\gamma^{-3/2}$ for $\gamma=1$, 1.1 and $10^3$,
respectively, all with $\theta_B=30^\circ$. The O-modes have two
branches: the superluminous branch ($w>ck$ or $n<1$) and the
subluminous branch ($w<ck$ or $n>1$), the latter corresponds to plasma
oscillations. At low densities, $(\omega_{\rm
p}/\omega)\gamma^{-3/2}\ll 1$, the superluminous O-mode becomes
transverse vacuum electromagnetic wave which can escape from the
magnetosphere.

In the very low density region, $(\omega_{\rm p}/\omega)
\gamma^{-3/2}\ll1$, the QED effect may not be neglected compared to
the plasma effect.  The ``competition'' between the vacuum
polarization effect and the plasma effect gives rise to a vacuum
resonance, at which the superluminous O-mode and the X-mode may be
coupled with each other. In the remainder of this paper, we will focus
on this vacuum resonance phenomenon.



\subsection{Wave Modes Including QED Effect}
We now consider how vacuum polarization affects the X-mode and
(superluminous) O-mode. Because of the possibility of ``mode
crossing'', we label the modes as ``+ mode'' and ``$-$ mode''. We write
the mode polarization vector as $\vecE_{\pm}=\vecE_{\pm T}+\vecE_{\pm
z}\vechatz$, with the transverse part
\begin{equation}
\vecE_{\pm T}=\frac{1}{(1+K_{\pm}^2)^{1/2}}(iK_{\pm},1),
\label{eq:eigenvector}
\end{equation}
where $iK_{\pm}=E_x/E_y$. Eliminating $n^2$ from
equation~(\ref{eq:nrefractmatrix}), we obtain
\begin{equation}
K_{\pm} = \polarb\pm\sqrt{\polarb^2+r} \simeq \polarb\pm\sqrt{\polarb^2+1},
 \label{eq:polark}
\end{equation}
where the polarization parameter $\polarb$ is given by
\begin{eqnarray}
\polarb & = & -i\frac{\vpr\eta_{xx}-\eta_{yy}}{2\eta_{yx}} \nonumber\\
 &=& -\frac{\lp S'^2-D^2-S'P'+C^2\rp\sin^2\thetab+\lp A^2-C^2\rp + 
     \lp AS'-DC\rp\sin2\thetab + \lp S'P'-A^2\rp\lp 1-r\rp}
     {2\lb P'D\cos\thetab-S'C\sin\thetab+A\lp D\sin\thetab-C\cos\thetab\rp \rb}.
\label{eq:polarb1}
\end{eqnarray}
The index of refraction $n_{\pm}$ of the two modes can be 
obtained from equation~(\ref{eq:nrefractmatrix}), giving
\begin{equation}
n_{\pm}^2 = \frac{\eta_{yy}}{r}+\frac{\eta_{yx}}{r}iK_{\pm}. \label{eq:n1}\\
\end{equation}

We now consider the case of cold streaming plasma with $\beta_-=\beta_+=\beta$.
We focus on parameter regimes satisfying the following conditions:
\begin{equation}
u\gamma^{-2}\ggthetanon^{-2} = 7.812\times10^{12}\,B_{12}^2\nuG^{-2}
 \gamma_3^{-2}\ggthetanon^{-2} \gg 1, \label{eq:condition1}
\end{equation}
\begin{equation}
v\gamma^{-1} =5.617\times10^{-3} \eta B_{12} \Ps^{-1} \nuG^{-2} 
 \gamma_3^{-1} \ll 1, \label{eq:condition2}
\end{equation}
where $\nu$ is the wave (photon) frequency, $\nuG=\nu/({\rm 1\,GHz})$ and
$\gamma_3=\gamma/10^3$.
The first condition implies that the Doppler-shifted frequency is lower
than $\wc$, so that we can use equation~(\ref{eq:fzetasimple}) for $f_{11}$,
$f_{12}$ and $f_{\eta}$. The second condition implies
$|f_{12}|^2\ll|f_{11}|\ll1$, and $|f_{\eta}|\ll1$. Under these
conditions, equation~(\ref{eq:polarb1}) reduces to
\begin{equation}
\polarb \simeq \frac{\lp f_{\eta}+q+m\rp\singhsq-f_{11}\lb1-\zetagh^2\rb}{2f_{12}\zetagh}.
\label{eq:polarb2}
\end{equation}
We shall see that conditions~(\ref{eq:condition1}) and
(\ref{eq:condition2}) also imply that the index of refraction is close
to unity, $|n-1|\ll1$. Thus $\zetagh^2 = (\cosgh-n\beta)^2 /
(1-n\beta\cosgh)^2 \approx 1-\gamma^{-2}\singhsq\ggthetanon^2$, and we
can easily check that the second term in the numerator of
equation~(\ref{eq:polarb2}) is much smaller than $f_{\eta}$.
Equation~(\ref{eq:polarb2}) therefore simplifies to
\begin{eqnarray}
\polarb &\simeq& \frac{f_{\eta}+q+m}{2f_{12}}\frac{\singhsq}{\zetagh} \nonumber\\
        &=& \frac{f_{\eta}}{2f_{12}}\frac{\singhsq}{\zetagh}
            \lp1+\frac{q+m}{f_{\eta}}\rp \nonumber\\
	&=&\polarbpl\polarbvp, \label{eq:polarb3}
\end{eqnarray}
where $\polarbpl$ is the polarization parameter in the absence of
vacuum polarization:
\begin{eqnarray}
\polarbpl  &\simeq& \frac{f_{\eta}}{2f_{12}}\frac{\singhsq}{\zetagh} \nonumber\\
       &\simeq& -\frac{1}{2}\lp1-2f\rp^{-1}u^{1/2}\gamma^{-3}\ggthetanon^{-3}
                \frac{\singhsq}{\zetagh}\label{eq:polarbpl},
\end{eqnarray}
and $\polarbvp$ is the correction factor due to vacuum polarization:
\begin{eqnarray}
\polarbvp &\simeq&  1 + \frac{q+m}{f_{\eta}} \nonumber \\
 &\simeq& 1-\frac{q+m}{v\gamma^{-3}\ggthetanon^{-2}}.
\label{eq:polarbV}
\end{eqnarray}

Equation~(\ref{eq:n1}) for the index of refraction $n_{\pm}$ of the two modes 
also simplifies to
\begin{equation}
n_{\pm}^2 \simeq 1-m\sin^2\thetab+f_{11}+f_{12}\lp\cos\thetab-\zeta \sin\thetab\rp K_{\pm}. 
 \label{eq:n2}
\end{equation}
For positive $\polarb$ with $|\polarb|\gg1$,
we obtain simple expressions of refractive indices
\begin{equation}
n_+^2=1-m\sin^2\thetab+f_{11}+(f_\eta+q+m)\sin^2\thetab, \qquad
n_-^2=1-m\sin^2\thetab+f_{11}, \label{eq:n3}
\end{equation}
while for negative $\polarb$ with $|\polarb|\gg1$, we have
\begin{equation}
n_+^2=1-m\sin^2\thetab+f_{11}, \qquad
n_-^2=1-m\sin^2\thetab+f_{11}+(f_\eta+q+m)\sin^2\thetab. \label{eq:n4}
\end{equation}
For $|\polarb|=0$, we have $K_{\pm}=\pm 1$ and
\begin{equation}
n_{\pm}^2=1-m\sin^2\thetab+f_{11}\pm f_{12}\lp\cos\thetab-\zeta \sin\thetab\rp . \label{eq:n5}
\end{equation}
From equations~(\ref{eq:n3}--\ref{eq:n5}), we see that $n$ is indeed close to unity 
when equations~(\ref{eq:condition1}) and (\ref{eq:condition2}) are satisfied.

\subsection{Vacuum Resonance}

For $|\polarb|\gg1$, the two modes are (almost) linearly polarized:
the mode with $|K|\simeq 2|\polarb|\gg1$ is polarized in the
$\vechatk$-$\vechatB$ plane, and is usually called ordinary mode
(O-mode); the mode with $|K|\simeq 1/(2|\polarb|)\ll1$ is polarized
perpendicular to the $\vechatk$-$\vechatB$ plane, and is called
extraordinary mode (X-mode). From equation~(\ref{eq:polarb3}), we see
that for a general $\thetab$ which is not too close to $0^o$ or
$180^o$, and for almost all values of $B$, $N$, $\nu$, $\gamma$'s, the
inequality $|\polarb|\gg1$ is satisfied either when the condition
\begin{eqnarray}
\left|\frac{f_{\eta}}{f_{12}}\sin^2\theta_B\right| 
&\simeq& (1-2f)^{-1}u^{1/2}\gamma^{-3}(1-\cos\theta_B)^{-3}\sin^2\theta_B \nonumber\\
&=& 2.795\,B_{12}\nuG^{-1}\gamma_3^{-3}(1-\cos\theta_B)^{-3}\sin^2\theta_B(1-2f)^{-1} \gg 1 
\label{eq:linear1}
\end{eqnarray}
is satisfied, or when
\begin{eqnarray}
\left| \frac{q+m}{f_{12}}\sin^2\theta_B\right| &\simeq& \frac{\alpha_F}{15\pi}
     \lp\frac{B}{B_Q}\rp^2\lp1-2f\rp^{-1}u^{1/2}v^{-1}(1-\cos\theta_B)^{-1}\sin^2\theta_B \nonumber\\
     &=& 39.56\,B_{12}^2\nuG\lp1-2f\rp^{-1}\eta^{-1}\Ps
     (1-\cos\theta_B)^{-1}\sin^2\theta_B \gg 1 \label{eq:linear2}
\end{eqnarray}
is satisfied. The exception occurs when 
\begin{equation}
f_{\eta}+q+m=0 \label{eq:vr_0}
\end{equation}
or 
\begin{equation}
\polarbvp = 1-\frac{q+m}{v\gamma^{-3}\ggthetanon^{-2}}=0. \label{eq:vr}
\end{equation}
This defines the ``vacuum resonance''. For given $\nu$, $\gamma$ and 
$B$, the resonance occurs at the density
\begin{equation}
N_V=9.905\times 10^{11} B_{12}^2\nuG^2\gamma_3^3
           \ggthetanon^2F(b)\,{\rm cm}^{-3},
\label{eqrhores}
\end{equation}
where 
\begin{equation}
F(b)\equiv {q+m \over \alpha_F^2 b^2/(15\pi)}
\end{equation}
is equal to unity for $b=B/B_Q\ll 1$ and is at most of order a few for
$B\lo 10^{15}$~G (see Fig.~1 of Ho \& Lai 2003).  It's obvious that
$N_V\propto\gamma^3$ for $\gamma\gg1$ and $\theta_B$ not too close to
$0^\circ$ (see Figure~\ref{fig:n-eta_g}). For $\gamma=1$,
equation~(\ref{eqrhores}) agrees with the result of Lai \& Ho~(2002).
The dependence of the resonance density $N_V$ on $\gamma$ and
$\theta_B$ can be easily understood from the cold (non-streaming)
plasma limit and Lorentz transform. The wave freqency in the plasma
rest frame is $\omega'=\gamma (\omega-\beta k_\parallel)\simeq \gamma
\omega (1-\beta\cos\theta_B)$. Note that in this frame, the external
magnetic field and the dielectric tensor due to vacuum polarization
are unchanged. The rest frame plasma density at the vacuum resonance
is $N_V'\propto B^2{\omega'}^2F(b)$, and the corresponding "lab frame"
density is $N_V=\gamma N_V'$. 
We can rewrite the resonance density in
terms of the Goldreich-Julian density
\begin{equation}
\eta_V = \frac{N_V}{\NGJ} = 14.15\Ps B_{12}\nuG^2\gamma_3^3
             \ggthetanon^2 F(b). \label{eq:etares}\\
\end{equation}
The physical meaning of the resonance is clear: For given $\nu$, $\gamma$
and $B$, the dielectric property of the medium is dominated by the plasma 
effect when $N\gg N_V$, while it is dominated by vacuum polarization
when $N\ll N_V$; at $N=N_V$, the plasma effect and vacuum polarization
compensate each other, and the wave modes become exactly circular polarized.

There are other ways to view the vacuum resonance. For example, at
given $B$, $\gamma$ and density $N$ (or $\eta$), we can define the
vacuum resonance frequency:
\begin{equation}
\nu_V = 0.266\lb \Ps^{-1}B_{12}^{-1}\eta
          \gamma_3^{-3}\ggthetanon^{-2}F^{-1}\rb^{1/2}\,{\rm GHz}. 
\label{eq:nuV}
\end{equation}
Thus, wave modes with $\nu \ll \nu_V$ are determined by the plasma
effect, while those with $\nu \gg \nu_V$ are determined by the vacuum
polarization effect.

The characteristic width of the resonance region 
can be estimated by considering $|\polarb|=1$ as defining
the edge of the resonance. Since 
$\beta_V=1-N_V/N$, we find that
the densities at the edges of the resonance are $N_V\pm\Delta N$, 
with
\begin{equation}
\frac{\Delta N}{N_V}\simeq \frac{1}{\left|\polarbpl\right|}
        =0.7157\lp 1-2f\rp B_{12}^{-1}\nuG\gamma_3^3\ggthetanon^3
	\frac{\lp\cos\thetab-\zeta \sin\thetab\rp}{\sin^2\thetab},
	\label{eq:reswidth}
\end{equation}
where $\cos\theta_B-\zeta\sin\theta_B\simeq (\cos\theta_B-\beta)/(1-\beta\cos
\theta)$.

\begin{figure}
\centering
\includegraphics[height=10cm, angle=-90]{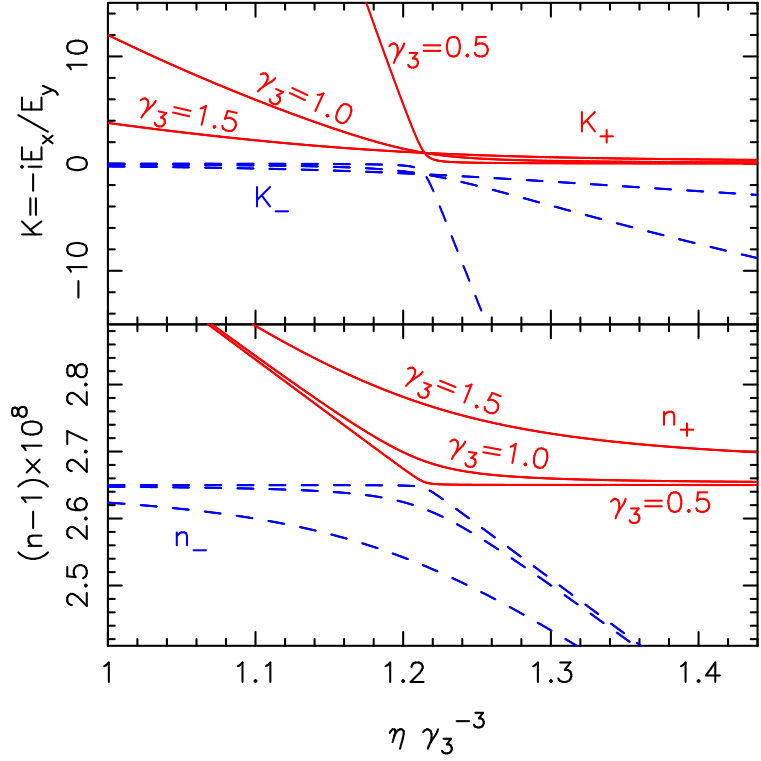}
\caption{ The polarization ellipticity $K$ (upper panel) and index of
refraction $n$ (lower panel) of the wave mode as a function of the
plasma density parameter $\eta=N/\NGJ$ (see equation~\ref{eq:NGJ})
near the vacuum resonance for $B=10^{12}\,{\rm G}$, $\nu=1\,{\rm
GHz}$, $f=0$ and $\theta_B=45^o$. Three values of $\gamma$ are
considered: $\gamma_3=\gamma/10^3=0.5$, 1, 1.5. The solid lines are
for the ``+'' mode and dashed lines for the ``$-$'' mode. Both the
resonant density (see Eq.~\ref{eq:etares}) and width (see
Eq.~\ref{eq:reswidth}) are proportional to $\gamma^3$
for $\gamma\gg 1$.
\label{fig:n-eta_g}
}
\end{figure}

\begin{figure}
\centering
\includegraphics[height=10cm, angle=-90]{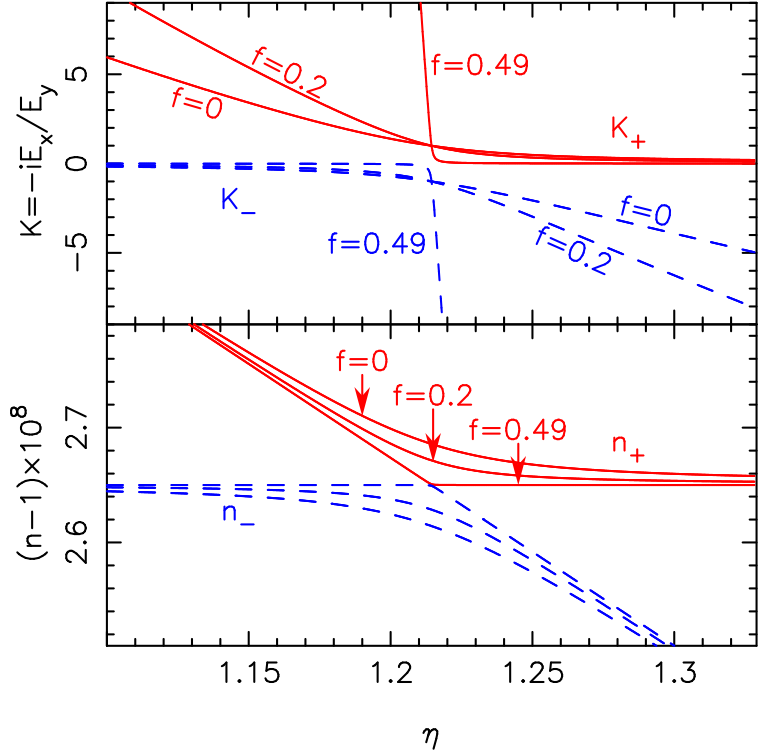}
\caption{ Same as Fig.~\ref{fig:n-eta_g}, except for $B=10^{12}\,{\rm
G}$, $\nu=1\,{\rm GHz}$, $\gamma=10^3$, $\thetab=45^o$ and different
values of the plasma positron fraction: $f=0$, 0.2 and 0.49.
\label{fig:n-eta_f}
}
\end{figure}

Figures~\ref{fig:n-eta_g} and \ref{fig:n-eta_f} show the mode
properties near vacuum resonance for different values of $\gamma$ and
$f$. It's obvious that the resonance density and width scale with
$\gamma^3$ (for $\gamma\gg1$), and the resonance density doesn't
change with $f$, while the resonance region becomes narrow when $f$ is
close to 0.5.

Note that while vacuum resonance can always be located by
$f_\eta+q+m=0$ [Eqs.~(\ref{eq:vr_0}) and (\ref{eq:vr})], significant
``avoided mode crossing'' (as illustrated in Figs.~\ref{fig:n-eta_g}
and \ref{fig:n-eta_f} and the inserts of
Figs.~\ref{fig:n_v}~--~\ref{fig:n_v_g1000}) occurs only if equations
(\ref{eq:linear1}) and (\ref{eq:linear2}) are satisfied. If these
``linear polarization'' conditions are not satisfied, the modes will
be approximately circular polarized even away from the resonance, and
no dramatic change in the mode properties takes place around the
vacuum resonance (see Fig.~\ref{fig:nonlinear}).

\begin{figure}
\centering
\includegraphics[height=10cm, angle=-90]{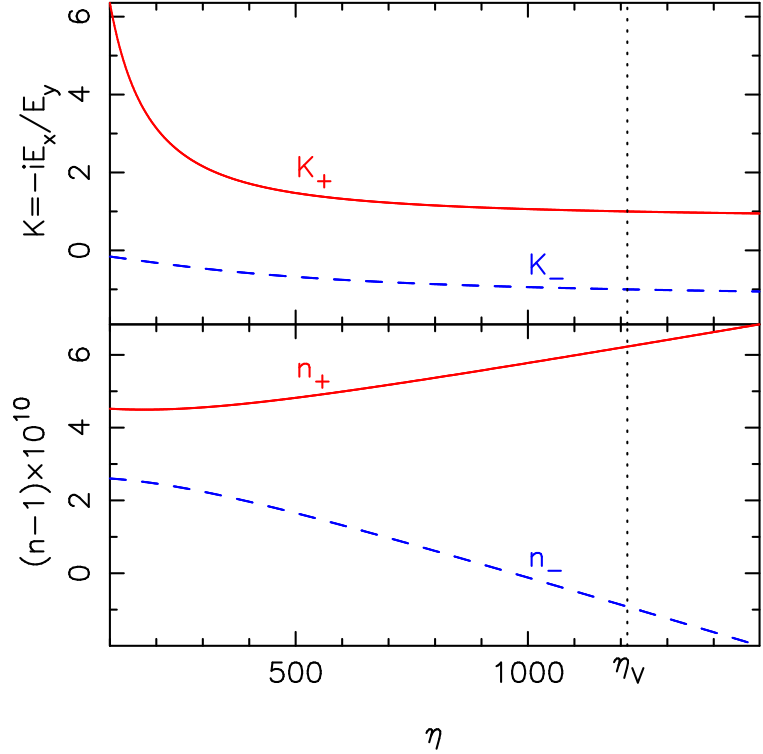}
\caption{ Same as Fig.~\ref{fig:n-eta_g}, except for $B=10^{11}\,{\rm
G}$, $\nu=100\,{\rm GHz}$, $1-2f=0.1$, $\gamma=10^3$, $\thetab=45^o$.
The vacuum resonance (defined by $f_\eta+q+m=0$) occurs at
$\eta=1.214\times10^3$ (the vertical dotted line). In this case,
equations~(\ref{eq:linear1}) and (\ref{eq:linear2}) are not satisfied,
and ``avoided mode crossing'' does not occur near the vacuum
resonance.
\label{fig:nonlinear}
}
\end{figure}

\section{Mode Evolution Across the Vacuum Resonance}

Consider a photon (or electromagnetic wave) of a given frequency $\nu$
and polarization state propagating in the NS magnetosphere. The
magnetosphere is inhomogeneous because of variations in $\vecB$, $N$
and possibly $\gamma$. How does the polarization of the photon
evolve, particularly as the photon traverses the vacuum resonance
region (e.g. from the plasma-dominated region to the vacuum-dominated
region)? Clearly, if the variations of the magnetosphere parameters
($\vecB$, $N$, etc.) are sufficiently gentle, the polarization state
of the photon will evolve adiabatically, i.e. a photon in a definite
wave mode will stay in that mode. Then Figure~\ref{fig:n-eta_g} and
\ref{fig:n-eta_f} show that across the vacuum resonance, the photon
polarization ellipse will rotate by $90^o$, with the mode helicity
unchanged.

To quantify the mode evolution, it is convenient to introduce the
``mixing'' angle $\theta_m$ via $\tan\theta_m=1/K_+$, so that
\begin{equation}
\tan 2\theta_m=\polarb^{-1},
\label{eq:thetam}\end{equation}
where we have used $|r-1|\ll 1$. The transverse eigenvectors of the
modes are $\vecE_{+T}=(i\cos\theta_m,\sin\theta_m)$ and
$\vecE_{-T}=(-i\sin\theta_m,\cos\theta_m)$.  Clearly, at the
resonance, $\theta_m=45^\circ$, the X-mode and O-mode are maximally
``mixed''.

A general polarized electromagnetic wave with frequency $\omega$
traveling in the $z$-direction can be written as a superposition of
the two modes:
\begin{equation}
\vecE(z)=A_+(z)\vecE_+(z)+A_-(z)\vecE_-(z),
\label{field}
\end{equation}
Note that both $A_{\pm}$ and $\vecE_{\pm}$ depend on $z$.
Substituting equation~(\ref{field}) into the wave equation
\begin{equation}
\nabla\times\lp\bmu^{-1}\cdot\nabla\times\vecE\rp
  =\frac{\omega^2}{c^2} \veceps\cdot\vecE, 
\end{equation}
we obtain the amplitude evolution equations (see Lai \& Ho 2002)
\begin{equation}
i\left(\begin{array}{c}A_+'\\ A_-'\end{array}\right)
\simeq \left(\begin{array}{cc}
-\Delta k/2 & i\theta_m'\\
-i\theta_m' & \Delta k/2
\end{array} \right)
\left(\begin{array}{c}A_+\\ A_-\end{array}\right),
\label{eqap2}\end{equation}
where $'$ stands for $d/dz$, $\Delta k=k_+-k_-$. In deriving
equation~(\ref{eqap2}), we have assumed that $\vecE_{\pm}(z)$ and
$A_{\pm}(z)$exp$\lp-i\int^z k_{\pm}dz\rp$ vary on a length scale much
larger than the photon wavelength, and we have used $k_+\simeq k_-$
and $|k_\pm'/k_\pm|\ll |k_\pm|$. Clearly, when $|\theta_m'|\ll |\Delta
k/2|$, or
\begin{equation}
\Gamma\equiv \left|{(n_+-n_-)\omega\over 2\theta_m'c}\right|\gg 1,
\label{eq:gamma1}\end{equation}
the polarization vector will evolve adiabatically (e.g., a photon in
the plus-mode will remain in the plus-mode). Using
equations~(\ref{eq:polarb3}) and (\ref{eq:thetam}), we find
\begin{equation}
\theta'_m = -\frac{1}{4}\sin^2 2\theta_m
           \frac{f_{\eta}}{f_{12}}\frac{\singhsq}{\zetagh}
            \frac{\polarb'}{\polarbpl}.
\end{equation}
The difference in refractive indices of the two modes is
\begin{equation}
n_+-n_-\simeq {f_{12}\zetagh\over\sin 2\theta_m}.
\end{equation}
Thus equation~(\ref{eq:gamma1}) becomes
\begin{equation}
\Gamma = \frac{2\omega H}{c} \lp1-2f\rp^2\frac{\gamma^3\wpl^2}
      {\wc^2\sin^3(2\theta_m)}\ggthetanon^4
      \lp\frac{\cosgh-\zeta\singh}{\singh}\rp^2
      \gg 1, \label{eq:gamma2}
\end{equation}
where $H\equiv |\polarbpl/\polarb'|$ specifies the length scale
of variation of $\polarb$ along the ray. Equation~(\ref{eq:gamma2})
gives the general condition for adiabatic mode evolution along the
photon path.

Clearly, the adiabatic condition (\ref{eq:gamma1}) or
(\ref{eq:gamma2}) is most easily violated at the resonance
($\theta_m=45^\circ$).  Evaluating equation~(\ref{eq:gamma2}) at
$N=N_V$ and using equation~(\ref{eq:vr}) to eliminate $\omega_p^2$, we
obtain
\begin{equation}
\Gamma_V=(\nu/\nu_{\rm ad})^3, \label{eq:Gamma3}
\end{equation}
with the ``adiabatic frequency''
\begin{equation}
\nu_{\rm ad}=6.081\,\lp1-2f\rp^{-2/3}\gamma_3^{-2}\ggthetanon^{-2}
           F^{-1/3}\lp\frac{\singh}{\cosgh-\zeta\singh}\rp^{2/3}
	   H_6^{-1/3}\,{\rm GHz}, \label{eq:nuad1}
\end{equation}
where we have used $H\simeq|dz/d\polarbvp|$ and $H_6=H/(10^6\,{\rm
cm})$. Since $\polarbvp=(1-N_V/N)$ and $\polarbpl$ changes very slowly
near resonance, for a constant $\vecB$, $\gamma$, the length
$H=N/|dN/dz|$ (evaluated at $N=N_V$) becomes the density scale height
along the ray.  For $\gamma=1$, equations~(\ref{eq:Gamma3}) and
(\ref{eq:nuad1}) agree with the result of Lai \& Ho~(2002). Using
equation~(\ref{eq:nuV}) to eliminate $\gamma$, we can also rewrite
equation (\ref{eq:nuad1}) as
\begin{equation}
\nu_{\rm ad} \simeq 35.59 \,\lp1-2f\rp^{-2/3}\nuG^{4/3}\eta^{-2/3}
      \lp \Ps B_{12}\rp^{2/3}F^{1/3}H_6^{-1/3}\ggthetanon^{-2/3}
      \lp\frac{\singh}{\cosgh-\zeta\singh}\rp^{2/3}\,{\rm GHz}. \label{eq:nuad2}
\end{equation}

\section{The Influence of Velocity Distribution}
\label{sec:vspread}

The results of Sections~3 and 4 are for cold streaming plasma with a
single Lorentz factor $\gamma$. For a general distribution function
$f(\gamma)$, equations~(\ref{eq:nrefract})~--~(\ref{eta_xy}) and
equations~(\ref{eq:eigenvector})~--~(\ref{eq:n1}) still apply. For the
simplest flat distribution of $\gamma$ [Eq.~(\ref{eq:fdistribution})],
the dielectric tensor functions are given by
equation~(\ref{eq:f11fd})~--~(\ref{eq:fetafd}) for $\gamma_{\rm
min}\gg 1$ and $\gamma_{\rm max}\omega (1-n\beta\cos\theta_B)
\ll\omega_c$. If we assume $v \gamma_{\rm min}^{-1}\ll 1$ [see
Eq.~(\ref{eq:condition2})], so that $n_\pm \simeq 1$, we find that the
vacuum resonance density, width and adiabatic frequency are given by:
\begin{eqnarray}
\eta_V &=& \frac{N_V}{n_{GJ}} = 14.15
             \Ps B_{12}\nuG^2\gamma_{c3}^3
	     \lp1-\Delta\gamma^2/\gamma_c^2\rp^{2}
             \ggthetanon^2F, \label{eq:etaV_dg}\\
\frac{\Delta N}{N_V}&=&0.7157\lp 1-2f\rp B_{12}^{-1}\gamma_{c3}^3
         \lp1-\Delta\gamma^2/\gamma_c^2\rp^{2}
         \ggthetanon^3
         \frac{\lp\cos\theta_B-\zeta \sin\theta_B\rp}{\sin^2\theta_B},\label{eq:dN_dg}\\
\nu_{\rm ad}&=&6.081\,\lp1-2f\rp^{-2/3}\gamma_3^{-2}\ggthetanon^{-2}
           F^{-1/3}\lp\frac{\singh}{\cosgh-\zeta\singh}\rp^{2/3}
	   H_6^{-1/3}\,{\rm GHz}. \label{eq:nuad_dg}
\end{eqnarray}
Figure~\ref{fig:n-eta_dg} shows the polarization ellipticities and the
refractive indices of the two wave modes in plasmas with different
$\Delta\gamma$'s. Thus, a broad distribution of $\gamma$'s does not
qualitatively affect the vacuum resonance behavior.
\begin{figure}
\centering
\includegraphics[height=10cm, angle=-90]{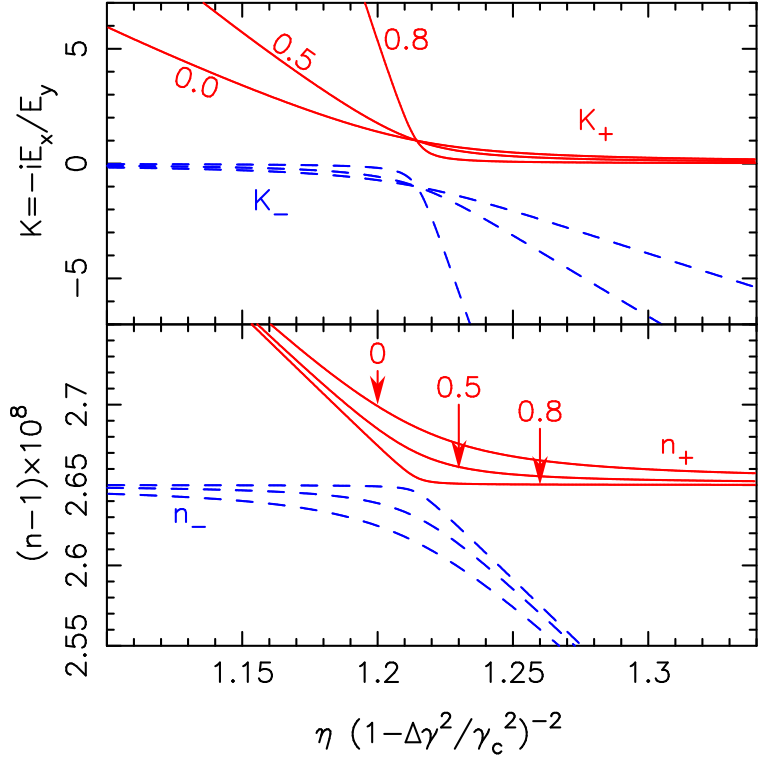}
\caption{ Same as Fig.~\ref{fig:n-eta_g}, except for plasmas with a flat
distribution of Lorentz factor (see
equation~\ref{eq:fdistribution}). The parameters are $B=10^{12}\,{\rm
G}$, $\nu=1\,{\rm GHz}$, $f=0$, $\gamma=10^3$, $\theta_B=45^o$.  For
each mode, the three lines are for $\Delta\gamma/\gamma_c=0$, 0.5,
0.8.
\label{fig:n-eta_dg}
}
\end{figure}

\section{Vacuum Resonance in a Plasma with Two Opposite Streams}
We now consider the case where the pair plasma is composed of two
opposite streams, for which the dielectric tensor is given in
Sec.2.1.2. The derivation of mode properties for this case are the
same as in Section 3. Similar to Eqs.~(\ref{eq:condition1}) and
(\ref{eq:condition2}), we assume $u\gamma^{-2}\lp1\pm
n\beta\cosgh\rp^{-2} \gg 1$ and $v\gamma^{-1}\ll 1$.  The polarization
parameter $\polarb$ (Eq.~\ref{eq:polarb1}) reduces to
\begin{equation}
\polarb \simeq \frac{\lp f_{\eta,+}+f_{\eta,-}+q+m\rp\singhsq
                +f_{11,+}\lb1-\lp\cosgh-\zeta_+\singh\rp^2\rb
                -f_{11,-}\lb1-\lp\cosgh-\zeta_-\singh\rp^2\rb}
	{2\lb f_{12,+}\lp\cosgh-\zeta_+\singh\rp
	     +f_{12,-}\lp\cosgh-\zeta_-\singh\rp \rb}. \label{eq:polarb_opp}
\end{equation}
where we have neglected the terms proportional to $f_{12, \pm}^2$. We
can see that Eq.~(\ref{eq:polarb_opp}) is rather similar to
Eq.~(\ref{eq:polarb2}). Since
\begin{equation}
\cosgh-\zeta_{\pm}\singh = \frac{\cosgh\pm\beta}{1\pm\beta\cosgh}
                     = \pm1 \mp \frac{1}{2}\gamma^{-2}\singhsq\lp1\pm\beta\cosgh\rp^{-2}, 
\end{equation}
the last two terms in the numerator of Eq.~(\ref{eq:polarb_opp}) can
be neglected compared to the first term. The denominator can be
simplified to $f_{12,+}-f_{12,-}\simeq vu^{-1/2}(1-M\beta\cosgh)$,
with $M\equiv1-2f$. Thus, we can rewrite Eq.~(\ref{eq:polarb_opp}) as
\begin{equation}
\polarb \simeq \frac{\lp f_{\eta}+q+m\rp\singhsq}
	{2vu^{-1/2}(1-M\beta\cosgh)}, \label{eq:polarb_opp2}
\end{equation}
with
\begin{equation}
f_{\eta}=f_{\eta,+}+f_{\eta,-}\simeq -v\gamma^{-3}
\frac{F_{\theta}}{\sin^4\thetab},
\end{equation}
and
\begin{equation}
F_{\theta}=1+\beta^2\cosghsq+2M\beta\cosgh,
\end{equation}
where we have assumed $\beta\simeq1$, $n\simeq1$ and
$\sin\theta_B\neq0$.  Similar to Sec.3.3, we write
$\polarb=\polarbpl\polarbvp$, where $\polarbpl$ is the polarization
parameter in the absence of vacuum polarization, and $\polarbvp$ is
the correction factor due to vacuum polarization:
\begin{eqnarray}
\polarbpl &\simeq& \frac{f_{\eta}\singhsq}{2vu^{-1/2}\lp1-M\beta\cosgh\rp} \nonumber\\
          &\simeq&  -\frac{1}{2}u^{1/2}\gamma^{-3}
                    \frac{F_{\theta}}
			 {\singhsq\lp1-M\beta\cosgh\rp}, \nonumber\\
\polarbvp &\simeq&  1+\frac{q+m}{f_{\eta}} \nonumber\\
          &\simeq&  1-\frac{q+m}{v\gamma^{-3}}
                    \frac{\sin^4\thetab}{F_{\theta}}.
\end{eqnarray}
Vacuum resonance occurs at $\polarbvp=0$, corresponding to the plasma
density (relative to the Goldreich-Julian density)
\begin{equation}
\eta_V = \frac{N_V}{\NGJ} = 14.15\Ps B_{12}\nuG^2\gamma_3^3
              \frac{\sin^4\thetab}{F_{\theta}}F(b). \label{eq:etares_opp}\\
\end{equation}
Similar to Eq.~(\ref{eq:reswidth}) for the single stream case, the characteristic width
of the resonance region is
\begin{equation}
\frac{\Delta N}{N_V}
        =0.7157B_{12}^{-1}\nuG\gamma_3^3
         \frac{\singhsq\lp1-M\beta\cosgh\rp}{F_{\theta}}.
	\label{eq:reswidth_opp}
\end{equation}
Similar to Eq.~(\ref{eq:gamma2}), the adiabatic condition is
\begin{equation}
\Gamma = \frac{2\omega H}{c} \frac{\gamma^3\wpl^2}
      {\wc^2\sin^3(2\theta_m)}
      \frac{\lp1-M\beta\cosgh\rp^2\singhsq}{F_{\theta}}
      \gg 1. \label{eq:gamma_opp2}
\end{equation}
At the resonance, we have $\Gamma_V=(\nu/\nu_{\rm ad})^3$, with
\begin{equation}
\nu_{\rm ad}=6.081\,\gamma_3^{-2}F^{-1/3}
             \lb\frac{F_{\theta}}
		     {\lp1-M\beta\cosgh\rp\sin^3\thetab}\rb^{2/3}
	   H_6^{-1/3}\,{\rm GHz}, \label{eq:nuad1_opp}
\end{equation}

\begin{figure}
\centering
\includegraphics[height=10cm, angle=-90]{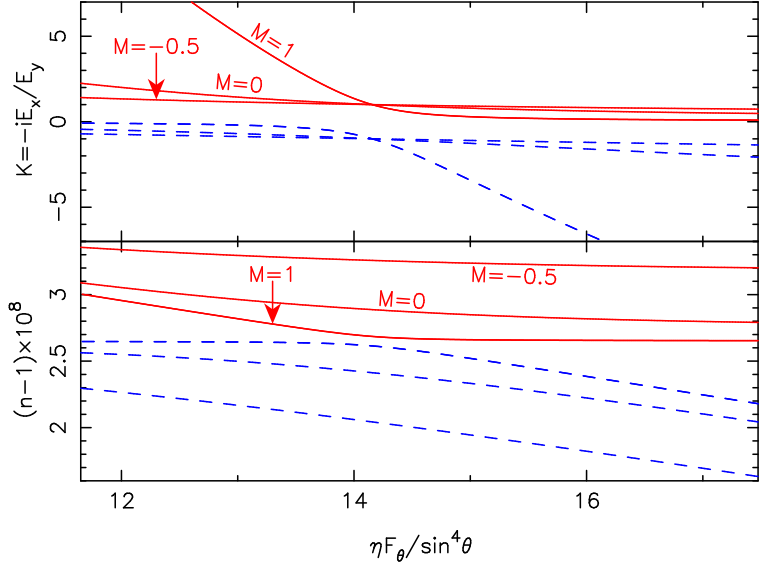}
\caption{ Same as Fig.~\ref{fig:n-eta_g}, except for plasmas with two
opposite streams. Note the $x$-axis is $\eta
F_{\theta}/\sin^4\theta_B$.  The parameters are $B=10^{12}\,{\rm G}$,
$\nu=1\,{\rm GHz}$, $\gamma=10^3$, $\theta_B=45^\circ$.  For each
mode, the three lines are for $M=1-2f=1$, 0, $-0.5$, respectively.
\label{fig:n-eta_opp}
}
\end{figure}
Figure \ref{fig:n-eta_opp} shows the mode evolution
near the vacuum resonance for different values of $M$'s (or
$f$'s). For the pair plasma with $M=0$, vacuum resonance still occurs
while the resonance region is much wider compared with $M>0$. For
$-1<M<0$, the modes evolution behaves as the case in
Fig.~\ref{fig:nonlinear}, since the linear condition is not satisfied in
this parameter regime (given in the caption of the figure).

If there is a velocity spread of the electrons and positrons, for
example the flat distribution given by Eq.~(\ref{eq:fdistribution}),
the equations above just need to be modified by appropriate factors
similar to Eqs.~(\ref{eq:etaV_dg})~--~(\ref{eq:nuad_dg}):
\begin{eqnarray}
\eta_V &=& \frac{N_V}{\NGJ} = 14.15\Ps B_{12}\nuG^2\gamma_3^3
              \lp1-\Delta\gamma_c^2/\gamma^2\rp^2
              \frac{\sin^4\thetab}{F_{\theta}}F(b). \\
\frac{\Delta N}{N_V}
       &=& 0.7157B_{12}^{-1}\nuG\gamma_3^3\lp1-\Delta\gamma_c^2/\gamma^2\rp^2
         \frac{\singhsq\lp1-M\beta\cosgh\rp}{F_{\theta}}.\\
\nu_{\rm ad} &=& 6.081\,\gamma_3^{-2}F(b)^{-1/3}
             \lb\frac{F_{\theta}}
		     {\lp1-M\beta\cosgh\rp\sin^3\thetab}\rb^{2/3}
	   H_6^{-1/3}\,{\rm GHz}.
\end{eqnarray}

\section{Discussion}

In previous sections, we have studied the property of wave propagation
in the magnetospheres of pulsars or magnetars for various plasma
parameters.  We have focused the vacuum resonance phenomenon, arising
from the combined effects of plasma and vacuum polarization. The
possible occurrence of the vacuum resonance and the related wave
property depends on the plasma parameters, magnetic field and the wave
frequency. The key equations are (assuming single-stream plasma):\\
(i) The vacuum resonance condition, Eq.~(\ref{eqrhores}) or
(\ref{eq:etares});\\
(ii) The adiabatic condition,
Eqs.~(\ref{eq:gamma2}~--~\ref{eq:nuad2});\\
In deriving the analytical expressions for the vacuum resonance, we
have assumed\\
(iii) The Doppler-shifted wave frequency is much less than the
electron cyclotron frequency, i.e., $\omega'=\gamma \omega
(1-n\beta\cosgh)\ll\omega_c$ [Eq.~(\ref{eq:condition1})];\\
(iv) The plasma is weakly dispersive, i.e., $v\gamma^{-1}\ll 1$
[Eq.~(\ref{eq:condition2})].\\ 
The vacuum resonance is particularly interesting in the parameter
regime such that either\\
(v) the waves are linearly polarized away from the vacuum resonance 
due to the plasma effect
[Eq.~(\ref{eq:linear1})], or\\
(vi) the waves are linearly polarized away from the vacuum resonance 
due to the vacuum polarization
[Eq.~(\ref{eq:linear2})].\\ 
Note that at the vacuum resonance, the waves are always circular polarized.

Figures \ref{fig:Bnu1} and \ref{fig:Bnu2} summarize these conditions
for two different sets of parameters.  In both figures, we see that
when the vacuum resonance induces significant ``avoided mode
crossing'' (cf. Figs. \ref{fig:n-eta_g} and \ref{fig:n-eta_f}),
i.e. when the resonance lies above the ``Linear I'' or ``Linear II''
line, wave evolution across the vacuum resonance is nonadiabatic. In
general, this can be understood as follows. We define the cross
frequency of the ``Linear I'' line and ``Linear II'' line (as well as
``Vacuum Resonance'' line) as $v_{\rm cross}$. We find
\begin{equation}
\nu_{\rm cross} \simeq
  0.58(1-2f)^{-1/3}\eta^{1/3}\sin\theta_B^{2/3}P_1^{-1/3}
  \gamma_3^{-2}(1-\cos\theta_B)^{-5/3}\,{\rm GHz}.
\end{equation}
Comparing to the adiabatic frequency $\nu_{\rm ad}$, we have
\begin{equation}
\frac{\nu_{\rm cross}}{\nu_{\rm ad}} \simeq
  0.1(1-2f)^{1/3}\eta^{1/3}P_1^{-1/3}
  (1-\cos\theta_B)^{1/3}F^{1/3}H_6^{1/3}.
\end{equation}
For typical parameters(e.g., $f=0-0.5$, $\eta=10^2-10^3$, $P\sim$1s,
$\theta=45^\circ$, $H_6=1$), $\nu_{\rm cross}$ is less than the
adiabatic frequency. Adiabatic mode evolution across the vacuum
resonance with appreciable mode crossing is possible for larger $\eta$
and $H_6$.

We now discuss possible implications of our results for various radiation
processes in pulsars and magnetars. We assume 
a dipole magnetic field, with
\begin{equation}
B\approx B_{\ast}\lp\frac{R_{\ast}}{r}\rp^3, \label{eq:Bsurface}
\end{equation}
where $B_{\ast}$ is the surface magnetic field and $R_{\ast}$ the
radius of the NS star. 
Substituting equation~(\ref{eq:Bsurface}) into
equation~(\ref{eq:etares}), we obtain the location of vacuum resonance
(assuming constant $\eta$ and $\gamma$)
\begin{equation}
\frac{r_V}{R_{\ast}} \simeq 0.5\,\lp\frac{\nu}{1\,{\rm GHz}}\rp^{2/3}X
       \lp\frac{B_{\ast}}{10^{12}\,{\rm G}}\rp^{1/3}
       \lp\frac{\gamma}{10^3}\rp\lp\frac{\eta}{10^2}\rp^{-1/3}
       \lp\frac{P}{1\,{\rm s}}\rp^{1/3}F^{1/3}\ggthetanon^{2/3}. \label{eq:resr}
\end{equation}
Since the dispersion due to vacuum polarization is of order $q+m
\propto F(b)B^2\propto r^{-6}$, while the plasma effect is measured by
$\sim v\gamma^{-3}\propto N\gamma^{-3}\propto \eta B\gamma^{-3}
\propto \eta\gamma^{-3} r^{-3}$, if $\eta\gamma^{-3}$ does not vary
rapidly, we find that for a given photon frequency $\nu$, the wave
dispersion is dominated by the vacuum effect for $r\lo r_V$ and
by the plasma effect for $r\go r_V$.

First consider the radio emission from the open field line region of a
pulsar.  The emission angle relative the local magnetic field line is
$\theta_B\sim1/\gamma$, so that $1-\beta\cosgh \simeq \gamma^{-2}$.
This would imply $r_V/R_\ast\ll 1$, even for $B_\ast\sim 10^{15}$~G
and for high frequencies (e.g. $\nu=20$~GHz). Along the ray
trajectory, the angle $\theta_B$ increases. In the small angle
approximation ($\theta_B\ll 1$), we have
\begin{equation}
\theta_B\approx \frac{3}{4}\sqrt{\frac{r_{\rm em}}{R_\ast}}
\theta_0\lp1-\frac{r_{\rm em}}{r}\rp,
\label{eq:theta_B}
\end{equation}
where $r_{\rm em}$ is the radius of the emission point, and $\theta_0$
is the polar angle at the stellar surface of the emission field line.
Thus $\theta_B$ increases from $0$ (at $r=r_{\rm em}$) to
$(3/4)(r_{\rm em}/R_\ast)\theta_0$. As an example, for $r_{\rm
em}=2R_\ast$ and $\theta_0\sim \sqrt{R_\ast/R_{\rm LC}}\simeq 0.0145
P_{1}^{-1/2}$, equation~(\ref{eq:theta_B}) implies $\theta_B\lo 0.015
P_{1}^{-1/2}$. From Eq.~(\ref{eq:resr}), we find
\begin{equation}
\frac{r_V}{R_{\ast}} \lo 0.1\,\lp\frac{\nu}{20\,{\rm GHz}}\rp^{2/3}
       \lp\frac{B_{\ast}}{10^{15}\,{\rm G}}\rp^{1/3}
       \lp\frac{\gamma}{10^3}\rp^{-1/3}  \lp\frac{\eta}{10^2}\rp^{-1/3}
       \lp\frac{P}{1\,{\rm s}}\rp^{-1/3}F^{1/3}. 
\end{equation}
This means that for radio emission along open, dipole field lines, 
plasma effects always dominate the property of wave propagation and
vacuum resonance will not occur.

Radio emission may also come from the large-curvature magnetic field structure
(e.g., field lines with curvature radius $\sim R_\ast$). 
In this case, even if $\theta_B\ll 1$ at emission, it will 
become significantly large ($\sim45^\circ$) after the wave 
propagates a short distance of order $R_\ast$. Thus, according to 
Eq.~(\ref{eq:resr}), vacuum resonance can occur for sufficiently 
high frequencies and strong surface magnetic fields. This could be
the case with the high-frequency radio emission from the transient
AXP XTE J1810-197 (Camilo et al. 2006).

Finally, optical/IR radiation emitted from the neutron star surface or
near vicinity may experience the vacuum resonance while propagating
through the magnetosphere. The polarization of such radiation may probe
the physical conditions of the magnetosphere.

\begin{figure}
\centering
\includegraphics[height=10cm, angle=-90]{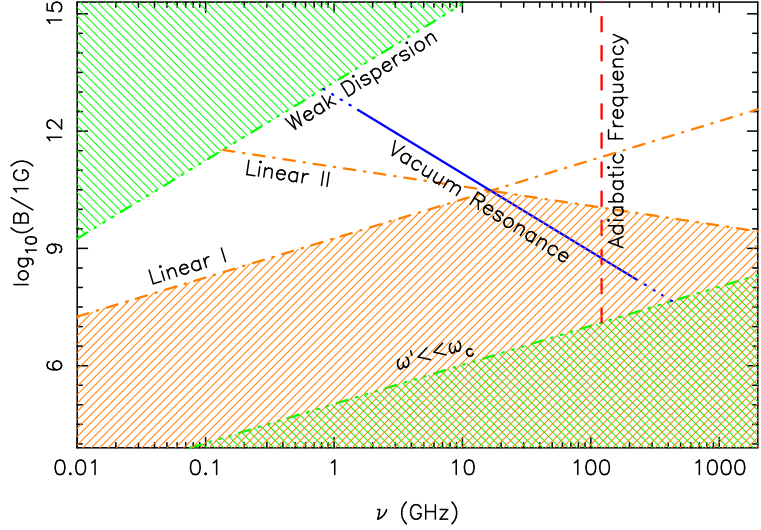}
\caption{ Vacuum resonance and wave properties in the magnetic field
  -- frequency domain for the plasma parameters $\eta=10$, $1-2f=0.1$,
  $\gamma_3=1$, $\thetab=45^o$, $H_6=10$ and the NS spin period
  $P=1\,{\rm s}$. The line labeled ``Vacuum Resonance'' gives the
  vacuum resonance condition [Eq.~(\ref{eq:etares})], the line labeled
  ``Adiabatic Frequency'' [Eq.~(\ref{eq:nuad1})] means that the
  adiabatic condition is satisfied to the right of the line when the
  wave evolves across the resonance. The line labeled
  ``$\omega'\ll\omega_c$'' [Eq.~(\ref{eq:condition1})] means that the
  Doppler-shifted frequency is much less than the electron cyclotron
  frequency above the line, and the line labeled ``weak dispersion''
  [Eq.~(\ref{eq:condition2})] means that the medium is weakly
  dispersive (i.e., index of refraction close to unity) below the
  line; Our analytical expression for the wave modes and vacuum
  resonance are valid in this parameter regime (below the ``weak
  dispersion'' line and above the ``$\omega'\ll\omega_c$''
  line). Above the ``Linear I'' line [Eq.~(\ref{eq:linear1})] or the
  ``Linear II'' line [Eq.~(\ref{eq:linear2})], the wave modes are
  linearly polarized except near the vacuum resonance.
\label{fig:Bnu1}
}
\end{figure}

\begin{figure}
\centering
\includegraphics[height=10cm, angle=-90]{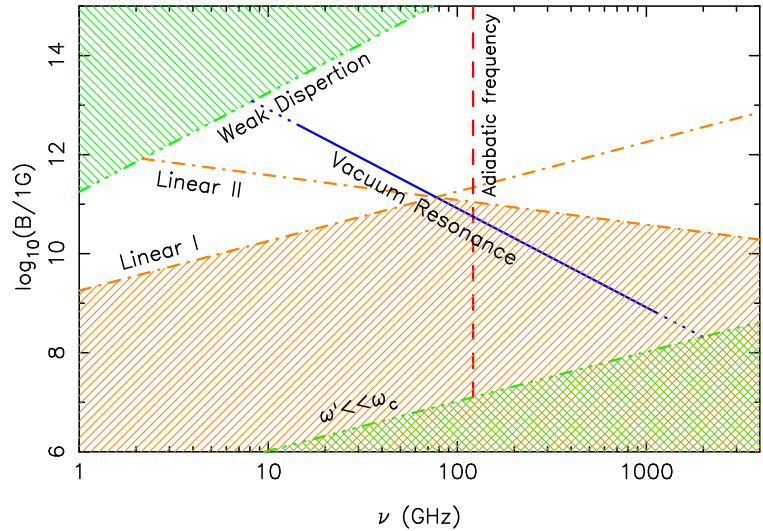}
\caption{Similar to Fig.~\ref{fig:Bnu1}, except with different
parameters: $\eta=1000$, $1-2f=0.1$, $\gamma_3=1$, $\thetab=45^o$,
$H_6=10$, $P=1\,{\rm s}$.
\label{fig:Bnu2}
}
\end{figure}

\section*{Acknowledgments}

We thank Qinghuan Luo for useful comment on an earlier draft of this
paper. This work is supported by National Natural Science Foundation
of China (10328305, 10473015 and 10521001). D.L. has also been
supported in part by NSF grant AST 0307252 and NASA grant NAG
5-12034. C.W. thanks the Astronomy Department at Cornell and
D.L. thanks NAOC (Beijing) for hospitality during the course of the
work.

\label{lastpage}


\begin{thebibliography}{longestkeymustbeshorterthanthis99}

\bibitem[]{} Adler, S. L. 1971, Ann. Phys., 67, 599


\bibitem[]{} Arons, J. \& Barnard, J. J. 1986, ApJ, 302, 120

\bibitem[]{} Asseo, E. \& Riazuelo, A. 2000, MNRAS, 318, 983

\bibitem[]{} Barnard, J. J. 1986, ApJ, 303, 280

\bibitem[]{} Barnard, J. J. \& Arons, J. 1986, ApJ, 302, 138

\bibitem[]{} Beloborodov, A. M. \& Thompson, C. 2006, astro-ph/0602417

\bibitem[]{} Blaskiewicz, M., Cordes, J. M., \& Wasserman, I. 1991, ApJ, 370, 643

\bibitem[]{} Camilo, F. et al. 2006, Nature, 442, 892

\bibitem[]{} Cheng, A. F. \& Ruderman, M. A., 1979, ApJ, 229, 348

\bibitem[]{} Daugherty, J. K. \& Harding, A. K. 1982, ApJ, 252, 337

\bibitem[]{} Gnedin, Yu. N., Pavlov, G. G., \& Shibanov, Yu. A. 1978, Sov. Astron. Lett., 4, 117

\bibitem[]{} Haberl, F. 2005, in 5 years of Science with XMM-Newton, MPE
   Report 288, 39­44 (astro-ph/0510480)

\bibitem[]{} Haberl, F., Motch, C., \& Zavlin, V. E. et al. 2004, A\&A, 424, 635

\bibitem[]{} Han, J. L., Manchester, R. N., Xu, R. X., \& Qiao, G. J. 1998, MNRAS, 300, 373

\bibitem[]{} Heisenberg, W. \& Euler, H. 1936, Z. Phys., 98, 714

\bibitem[]{} Heyl, J. S. \& Hernquist, L. 1997, Phys. Rev. D, 55, 2449

\bibitem[]{} Hibschman, J. A. \& Arons, J. 2001, ApJ, 554, 624

\bibitem[]{} Ho, W. C. G. \& Lai, D. 2003, MNRAS, 338, 233

\bibitem[]{} Kaplan, D. L., Kulkarni, S. R., \& van Kerkwijk, M. H. 2003, ApJ, 588, 33

\bibitem[]{} Kargaltsev, O. Y., Pavlov, G. G., Zavlin, V. E., \& Romani, R. W. 2005, ApJ, 625, 307

\bibitem[]{} Krall N. A. \& Trivelpiece A. W. 1986, Principles of Plasma Physics. San Francisco Press

\bibitem[]{} Kramer, M., Xilouris, K. M., Jessner, A., Lorimer, D. R., Wielebinski, R., \& Lyne, A. G., 1997, A\&A, 322, 846

\bibitem[]{} Kunzl, T., Lesch, H., Jessner, A. \& von Hoensbroech, A. 1998, ApJ, 505, 139

\bibitem[]{} Lai, D. \& Ho, W. C. G. 2002, ApJ, 566, 373

\bibitem[]{} Lai, D. \& Ho, W. C. G. 2003a, ApJ, 588, 962

\bibitem[]{} Lai, D. \& Ho, W. C. G. 2003b, PhRvL, 91, 1101L

\bibitem[]{} Lyubarskii, Y. E. \& Petrova, S. A. 1998, A\&A, 333, 181

\bibitem[]{} Lyutikov, M. 1998, MNRAS, 293, 447

\bibitem[]{} Melrose, D. B. 1973, Plasma Physics, 15, 99

\bibitem[]{} Melrose, D. B., Gedalin, M. E., Kennett, M. P., \& Fletcher, C. S. 1999, Journal of Plasma Physics 62, 233

\bibitem[]{} Melrose, D. B. \& Luo, Q. 2004, MNRAS, 352, 915

\bibitem[]{} Melrose, D. B. \& Stoneham, R. J. 1977, PASAu, 3, 120

\bibitem[]{} M\'{e}sz\'{a}ros, P. \& Ventura, J. 1979, Phys. Rev. D, 19, 3565

\bibitem[]{} Mignani, R. P. et al. 2006, astro-ph/0608025

\bibitem[]{} Mignani, R. P., de Luca, A., \& Caraveo, P. A. 2004, in ``Young 
Neutron Stars and Their Environments'', Proc. IAU Symp.218, Editors Camilo, F.,
Gaensler, B.M., ASP, p.391

\bibitem[]{} Petrova, L. I. 2006, MNRAS, 366, 1539

\bibitem[]{} Potekhin, A. Y., Lai, D., Chabrier, G. \& Ho, W. C. G. 2004, ApJ, 612, 1034

\bibitem[]{} Radhakrishnan, V. \& Rankin, Joanna M. 1990, ApJ, 352, 258

\bibitem[]{} Ruderman, M. 2003, astro-ph/0310777

\bibitem[]{} Schubert, C. 2001, in ``Quantum Electrodynamics and Physics of the Vacuum'',
Edited by Giovanni Cantatore, AIPC, V 564, p.28

\bibitem[]{} Stinebring, D. R., Cordes, J. M., Rankin, J. M., Weisberg, J. M., \& Boriakoff, V. 1984a, ApJS, 55, 247

\bibitem[]{} Stinebring, D. R., Cordes, J. M., Weisberg, J. M., Rankin, J. M., \& Boriakoff, V. 1984b, ApJS, 55, 279

\bibitem[]{} Thompson, C., Lyutikov, M., \& Kulkarni, S. R. 2002, ApJ, 574, 332

\bibitem[]{} Tsytovich, V. N. \& Kaplan, S. A. 1972, Astrofizika, 8, 441 
             (English transl. in Astrophysics, 8, 260 [1972])

\bibitem[]{} van Adelsberg, M. \& Lai, D. 2006, MNRAS, 373, 1495

\bibitem[]{} van Kerkwijk, M. H. \& Kaplan, D. L. 2006, astro-ph/0607320

\bibitem[]{} Wang, F. Y.-H., Ruderman, M., Halpern, J. P., \& Zhu, T. 1998, ApJ, 498, 373

\bibitem[]{} You, X. P. \& Han, J. L. 2006, ChJAA, 2, 237


\end{thebibliography}
\end{document}